\documentclass[onecolumn,12pt,epsfig]{IEEEtran}
\usepackage{ifpdf, flushend,subfigure}

\ifCLASSINFOpdf
  \usepackage[pdftex]{graphicx}
  \graphicspath{{../pdf/}{../jpeg/}}
  \DeclareGraphicsExtensions{.pdf,.jpeg,.png}
\else
  \usepackage[dvips]{graphicx}
  \graphicspath{{../eps/}}
  \DeclareGraphicsExtensions{.eps}
\fi
\usepackage[cmex10]{amsmath}
\usepackage {amssymb}
\usepackage{algorithmic}
\usepackage{array}
\usepackage{mdwmath}
\usepackage{mdwtab}
\usepackage{eqparbox}
\usepackage{algorithm}
\usepackage{algorithmic}

\renewcommand{\baselinestretch}{0.9}

\newcommand{\argmin}{\operatornamewithlimits{argmin}}
\newcommand{\argmax}{\operatornamewithlimits{argmax}}
\newcommand{\beq}{\begin{equation}}
\newcommand{\eeq}{\end{equation}}
\newcommand{\beqn}{\begin{eqnarray}}
\newcommand{\eeqn}{\end{eqnarray}}
\newcommand{\beqno}{\begin{eqnarray*}}
\newcommand{\eeqno}{\end{eqnarray*}}
\newcommand{\bma}{\begin{displaymath}}
\newcommand{\ema}{\end{displaymath}}
\newcommand{\bnu}{\begin{enumerate}}
\newcommand{\enu}{\end{enumerate}}
\newcommand{\bce}{\begin{center}}
\newcommand{\ece}{\end{center}}
\newcommand{\btb}{\begin{tabular}}
\newcommand{\etb}{\end{tabular}}

\begin{document}
%
\title{Channel Assignment with Access Contention Resolution for Cognitive Radio Networks}

\author{\IEEEauthorblockN{Le Thanh Tan and Long Bao Le}  
\thanks{The authors are with INRS-EMT, University of Quebec,  Montr\'{e}al, Qu\'{e}bec, Canada. 
Emails: \{lethanh,long.le\}@emt.inrs.ca. }}

\maketitle

\begin{abstract}
In this paper, we consider the channel allocation problem for throughput maximization in
cognitive radio networks with hardware-constrained secondary users. Specifically,
we assume that secondary users (SUs) exploit spectrum holes on a set of channels
where each SU can use at most one available channel for communication. 
\textbf{We present the optimal brute-force search algorithm and its complexity for this non-linear integer optimization problem.
Since the optimal solution has exponential complexity with the numbers of channels
and SUs, we develop two low-complexity channel assignment algorithms that can efficiently utilize spectrum
opportunities on these channels.} In the first algorithm, SUs are 
assigned distinct sets of channels. We show that this algorithm achieves the maximum throughput limit
if the number of channels is sufficiently large. In addition, we propose an overlapping
channel assignment algorithm, that can improve the throughput performance compared to
the non-overlapping channel assignment counterpart. In addition, we design a distributed
MAC protocol for access contention resolution and  integrate it into
the overlapping channel assignment algorithm. \textbf{We also analyze
the saturation throughput and the complexity of the proposed channel assignment algorithms.
Moreover, we have presented several potential extensions including greedy channel assignment 
algorithms under max-min fairness criterion and throughput analysis considering sensing errors}.  
Finally, numerical results are presented
to validate the developed theoretical results and illustrate the performance gains due to the proposed
channel assignment algorithms.

\end{abstract}

\begin{IEEEkeywords}
Channel assignment, MAC protocol, spectrum sensing, throughput maximization, cognitive radio.
\end{IEEEkeywords}

\section{Introduction}

Emerging broadband wireless applications have been demanding unprecedented increase in radio spectrum resources.
As a result, we have been facing a serious spectrum shortage problem. However, several recent
measurements reveal very low spectrum utilization in most useful frequency bands \cite{Zhao07}.
Cognitive radio technology is a promising technology that can fundamentally improve the
spectrum utilization of licensed frequency bands through secondary spectrum access.
However, transmissions from primary users (PUs) should be satisfactorily protected from 
secondary spectrum access due to their strictly higher access priority.

Protection of primary communications can be achieved through interference avoidance or interference
control approach (i.e., spectrum overlay or spectrum underlay) \cite{Zhao07}.
For the interference control approach, transmission powers of SUs
should be carefully controlled so that the aggregated interference they create
at primary receivers does not severely affect ongoing primary communications \cite{Le08}.
In most practical scenarios where direct coordination between PUs and SUs
is not possible and/or if distributed communications strategies are desired, it would be very
difficult to maintain these interference constraints. The interference avoidance approach
instead protects primary transmissions by requiring SUs to perform spectrum
sensing to discover spectrum holes over which they can transmit data \cite{R1}, \cite{Yu09}. 
This paper focuses on developing efficient channel assignment  
algorithms for a cognitive radio network with hardware-constrained secondary nodes
using the interference avoidance spectrum sharing approach. 

In particular, we consider the scenario where each SU can exploit at most one
available channel for communications. This can be the case if SUs
are equipped with only one radio employing a narrow-band RF front end \cite{So04}. In addition, it is 
assumed that white spaces are so dynamic that it is not affordable for each SU to sense all channels to discover
available ones and/or to exchange sensing results with one another. Under this setting, we
are interested in determining a set of channels allocated to each SU in advance
so that  maximum network throughput can be achieved in a distributed manner. To the best of
our knowledge, this important problem has not been considered before. The contributions of
this paper can be summarized as follows.

\begin{itemize}

\item 
We formulate the channel assignment problem for throughput maximization  as an integer optimization problem.
We then derive user and total network throughput for the case SUs are assigned distinct
sets of channels. \textbf{We present the optimal brute-force search algorithm and analyze its complexity}.

\item
We develop two greedy non-overlapping and overlapping channel assignment algorithms to solve the underlying NP-hard problem.
We prove that the proposed non-overlapping channel assignment algorithm achieves the maximum throughput as the number of
channels is sufficiently large. For the overlapping channel assignment algorithm, we design a MAC protocol for 
access contention resolution and we integrate the MAC protocol overhead analysis into the channel assignment algorithm.

\item
We analyze the saturation throughput and complexity of the proposed channel assignment algorithms. Moreover, we investigate
the impact of contention collisions on the developed throughput analytical framework.

\item
\textbf{We show how to extend the proposed channel assignment algorithms when max-min fairness is considered.
We also extend the throughput analytical model to consider sensing errors and propose an alternative MAC
protocol that can relieve congestion on the control channel}.

\item 
We demonstrate through numerical studies the interactions among various MAC protocol parameters and suggest its configuration.
We show that the overlapping channel assignment algorithm can achieve noticeable network throughput improvement
 compared to the non-overlapping counterpart. In addition, we present the throughput gains due to both proposed 
channel assignment algorithms compared to the round-robin algorithms, which do not exploit the heterogeneity in the
channel availability probabilities.

\end{itemize}

The remaining of this paper is organized as follows. In Section~\ref{Relworks}, we discuss important related works on spectrum sharing
algorithms and MAC protocols. Section~\ref{SystemModel} describes the system model and problem formulation. 
We present the non-overlapping channel assignment algorithm and describe its performance in Section IV.
The overlapping channel assignment and the corresponding MAC protocol are developed in Section V. Performance analysis of
the overlapping channel assignment algorithm and the MAC protocol is presented in Section VI. Several potential extensions
are discussed in Section VII. Section~\ref{Results} 
demonstrates numerical results followed by concluding remarks in Section~\ref{conclusion}.

\section{Related Works}
\label{Relworks}

Developing efficient spectrum sensing and access mechanisms for cognitive radio networks has been a very active 
research topic in the last several years 
\cite{R1}, \cite{Cor09}-\cite{chen07}. A great survey of recent works on MAC protocol design and analysis is given in \cite{Cor09}.
In \cite{R1}, it was shown that by optimizing the sensing time, a significant throughput gain
can be achieved for a SU. In \cite{Le11}, we extended the result in \cite{R1} to the multi-user setting where we design,
analyze, and optimize a MAC protocol to achieve optimal tradeoff between sensing time and contention overhead. In fact, we 
assumed that each SU can use all available channels simultaneously in \cite{Le11}. Therefore, the channel assignment problem
and the exploitation of multi-user diversity  do not exist in this setting, which is the topic of our current paper. Another related
effort along this line was conducted in \cite{Kim08} where sensing-period optimization and optimal channel-sequencing algorithms were
proposed to efficiently discover spectrum holes and to minimize the exploration delay.

In \cite{Su08}, a control-channel-based MAC protocol was proposed for secondary users to exploit white spaces in the
cognitive ad hoc network setting. In particular, the authors of this paper developed both random and negotiation-based
spectrum sensing schemes and performed throughput analysis for both saturation and non-saturation scenarios.
There exists several other synchronous cognitive MAC protocols, which rely on a control channel for spectrum negotiation and access
including those in \cite{Su07}, \cite{Nan07}, \cite{Le208}, \cite{Cor07}, \cite{Hsu07}.
A synchronous MAC protocols without using a control channel was proposed and studied in \cite{Konda08}.
In \cite{Do05}, a MAC layer framework was developed to dynamically reconfigure MAC and physical layer protocols.
Here, by monitoring current network metrics the proposed framework can achieve great performance by selecting the
best MAC protocol and its corresponding configuration. 

In \cite{Sala09}, a power-controlled MAC protocol was developed to efficiently exploit spectrum access opportunities
while satisfactorily protecting PUs by respecting interference constraints. Another
power control framework was described in \cite{shu06}, which aims to meet the rate requirements of SUs
and interference constraints of PUs. A novel clustering algorithm was devised in \cite{chen07} for
network formation, topology control, and exploitation of spectrum holes in a cognitive mesh network. It was shown that the proposed
clustering mechanism can efficiently adapt to the changes in the network and radio transmission environment.

\textbf{Optimal sensing and access design for cognitive radio networks were designed by using optimal stopping
theory in \cite{jia08}. In \cite{Sala10}, a multi-channel MAC protocol was proposed taking into account the distance
among users so that the white spaces can be efficiently exploited while satisfactorily protecting PUs.
Different power and spectrum allocation algorithms were devised to maximize the secondary network throughput in
\cite{taosu10}, \cite{wang11}, \cite{zhang113}. Optimization of spectrum sensing and access in which either cellular or TV bands can be
employed was performed in \cite{choi11}.}

\textbf{
In \cite{zhang11}, cooperative sequential spectrum sensing and packet scheduling
were designed for cognitive radios which are equipped with multiple spectrum sensors.
Energy-efficient MAC protocol was proposed for cognitive radio networks in \cite{zhang112}.
Spectrum sensing, access, and power control algorithms were developed  considering QoS protection 
for PUs and QoS provisioning for SUs in \cite{jeon12}, \cite{jha11}. Finally,
a channel hopping based MAC protocol was proposed in \cite{su08} for cognitive radio networks to alleviate the congestion
problem in the fixed control channel design. All these existing works, however, did not consider the scenario where
cognitive radios have hardware constraints which allows them to access at most one channel at any time. Moreover,
exploiting the multichannel diversity through efficient channel assignment is very critical to optimize the throughput
performance of the secondary network for this problem. We will investigate this problem considering its unique design issues in this paper.}

\section{System Model and Problem Formulation}
\label{SystemModel}


\subsection{System Model}
\label{System}

We consider a collocated cognitive radio network in which $M$ SUs exploit spectrum opportunities in  $N$ channels. We assume that 
any SU can hear the transmissions of other SUs. In addition, each SU can use at most one channel for 
its data transmission.
In addition, time is divided fixed-size cycle where SUs perform sensing on assigned channels
at the beginning of each cycle to explore available channels for communications.
 We assume that perfect sensing can be achieved with no sensing error. Extension to the imperfect
 spectrum sensing will be discussed in Section VII.B.
It is assumed that SUs transmit at a constant rate with the normalized value of one.

\subsection{Problem Formulation}
\label{ProbForm}

We are interested in performing channel assignment to maximize the system throughput. Let $T_i$
denote the throughput achieved by SU $i$. 
Let $x_{ij}$ describe the channel assignment decision where $x_{ij}=1$ if channel $j$ is assigned to SU
 $i$ and $x_{ij}=0$, otherwise. The throughput maximization problem can be formally written as follows:
\vspace{0.0cm}
\beqn
\label{Tput}
\max\limits_\textbf{x} \sum\limits_{i = 1}^M {{T_i}}.  \label{obj1} 
\eeqn
For non-overlapping channel assignments, we have following constraints
\beqn
 \sum\limits_{i = 1}^M {{x_{ij}} = 1}, \quad  \mbox{for\:all}\: j  \label{con1}.
\eeqn
We can derive the throughput achieved by SU $i$ for non-overlapping channel assignment as follows.
Let $S_i$ be the set of channels solely assigned to SU $i$. Let $p_{ij}$ be the probability that 
channel $j$ is available at SU $i$. For simplicity, we assume that $p_{ij}$ are independent from one another.
This assumption holds when each SU impacts different set of PUs on each channel. This can indeed be the case 
because spectrum holes depend on space. Note, however, that 
this assumption can be relaxed if the dependence structure of these probabilities is available. Under this assumption, $T_i$ can be calculated as
\beq \label{tput1}
T_i = 1 - \prod_{j \in \mathcal{S}_i} \overline{p}_{ij} = 1 - \prod\limits_{j = 1}^N {{{\left( {{{\bar p}_{ij}}} \right)}^{{x_{ij}}}}}
\eeq
where $\overline{p}_{ij} = 1 - {p}_{ij}$ is the probability that channel $j$ is not available for 
SU $i$. In fact,  $1 - \prod_{j \in \mathcal{S}_i} \overline{p}_{ij}$ is the probability that
there is at least one channel available for SU $i$. Because each SU can use
at most one available channel, its maximum throughput is 1. 
In the overlapping channel assignment scheme, constraints in (\ref{con1}) are not needed. 
From this calculation, it can be observed that the optimization problem (\ref{obj1})-(\ref{con1})
is a non-linear integer program, which is a NP-hard problem (\textbf{interest readers can refer to Part VIII of reference
\cite{Lee12} for detailed treatment of this hardness result}).

\subsection{Optimal Algorithm and Its Complexity}

\textbf{
Due to the non-linear and combinatorial structure of the formulated channel assignment problem, it would be impossible
to explicitly determine its optimal closed form solution. However, we can employ the brute-force search (i.e., exhaustive search) to determine the best channel assignment that results in the maximum total throughput. Specifically, we can enumerate all possible channel assignment solutions then determine
the best one by comparing their achieved throughput. This solution method requires a throughput analytical model
 that calculates the throughput for any particular channel assignment solution. We will develop such a model in Section \ref{tputana} of this paper.}

\textbf{
We now quantify the complexity of the optimal brute-force search algorithm, which is involved in determining all potential channel assignments.
Let us consider SU $i$ (i.e., $i \in \left\{1, \ldots, M\right\}$). Suppose we assign it $k$ channels where $k \in \left\{1, \ldots, N\right\}$). Then, there are $C_N^k$ ways to do so. Since $k$ can take any values in $k \in \left\{1, \ldots, N\right\}$, the total number of ways
to assign channels to SU $i$ is $\sum \limits_{k = 0}^N C_N^k = 2^N$. Hence, the total number of ways to assign channels to all SUs is $\left(2^N\right)^M = 2^{NM}$. Recall that 
we need to calculate the throughputs achieved by $M$ SUs for each potential assignment to determine the best one.  Therefore, the complexity of the optimal brute-force search algorithm is $O(2^{NM})$.}

\textbf{
Given the exponentially large complexity required to find the optimal channel assignment solution, we will develop 
sub-optimal and low-complexity channel assignment algorithms in the following sections. In particular, we consider two different 
channel assignment schemes. In the first scheme, SUs
are assigned distinct sets of channels. This channel assignment scheme simplifies the spectrum sharing
design because SUs do not compete for the same available channels. However, it overlooks
the potential diversity gain of the spectrum sharing problem.
In the second scheme, we allow SUs to sense and operate on overlapping channels. 
When one particular channel is exploited by several secondary users, it is assumed that
a MAC protocol is employed to resolve the channel contention.}

\section{Non-overlapping Channel Assignment Algorithm}
\label{nonover}

We develop a low-complexity  algorithm for non-overlapping channel assignment in this section. 
Recall that $\mathcal{S}_i$ is the set of channels solely assigned for SU
 $i$ (i.e., $\mathcal{S}_i \cap \mathcal{S}_j = \emptyset, \: i \neq j$). The greedy
channel assignment algorithm iteratively allocates channels to SUs that achieves the maximum increase
in the throughput. Detailed description of the proposed algorithm is presented in Algorithm 1.
In each channel allocation iteration, each SU $i$ calculates its increase in throughput
if the best available channel (i.e., channel  $j_i^* = \mathop {\arg \max }\limits_{j \in {\mathcal{S}_a}} \: {p_{ij}}$)
is allocated. This increase in throughput can be calculated as follows:
\beqn
\label{Tlem1}
 \Delta {T_i} = T_i^a - T_i^b = \left[ {1 - \left( {1 - {p_{ij_i^*}}} \right)\prod\limits_{j \in \mathcal{S}_i } {(1 - {p_{ij}})} } \right] 
 - \left[ {1 - \prod\limits_{j \in \mathcal{S}_i } {(1 - {p_{ij}})} } \right] 
= {p_{ij_i^*}} \prod\limits_{j \in \mathcal{S}_i } {(1 - {p_{ij}})}.  
\eeqn
It can be observed from (\ref{Tlem1}) that $\Delta {T_i}$ will quickly decrease over allocation iterations
because $\prod\limits_{j \in \mathcal{S}_i } {(1 - {p_{ij}})}$ tends to zero as the set $\mathcal{S}_i$ is expanded. We have the
following property for the resulting channel assignment due to Algorithm 1.

\begin{algorithm}[h]
\caption{\textsc{Non-Overlapping Channel Assignment}}
\label{mainalg}
\begin{algorithmic}[1]

\STATE Initialize the set of available channels  ${\mathcal{S}_a} := \left\{ {1,2, \ldots ,N} \right\}$ and $\mathcal{S}_i := \emptyset$ for $i=1, 2,\ldots , M$

\FOR{$i = 1$ to $M$}

\STATE $j_i^* = \mathop {\argmax }\limits_{j \in {\mathcal{S}_a}} \: {p_{ij}}$

\IF {$\mathcal{S}_{i} \neq 0$}

\STATE Find $\Delta {T_i} = T_i^a - T_i^b$,
where $T_i^a$ and $T_i^b$ is the throughputs after and before assigning channel $j_i^*$.

\ELSE

\STATE Find $\Delta {T_i} = p_{ij_i^*}$,

\ENDIF

\ENDFOR


\STATE ${i^*} = \argmax_i \Delta {T_i} $.

\STATE Assign channel $j_{i^*}^*$ to user $i^*$.

\STATE Update $\mathcal{S}_a = \mathcal{S}_a\backslash j_{i^*}^*$.

\STATE If $\mathcal{S}_a $ is empty, terminate the algorithm. Otherwise, return to step 2. 
\end{algorithmic}
\end{algorithm}

\vspace{0.1cm}
\noindent
\textbf{Proposition 1:} If we have $N >> M$, then the throughput achieved by any SU $i$ due to Algorithm 1 is very close
to the maximum value of 1.

\begin{proof}
This proposition can be proved by showing that if the number of channels is much larger than the number of
SUs (i.e., $N >> M$) then each SU will be assigned a large number of channels.
Recall that Algorithm 1 assigns channels to a particular SU $i$ based on the increase-in-throughput metric $\Delta {T_i}$.
This property can be proved by observing that if a particular SU $i$ has been assigned a large
number of channels, its $\Delta {T_i}$ is very close to zero. Therefore, other SUs who have been assigned a small
number of channels will have a good chance to receive more channels. As a result, all SUs are assigned a large
number of channels if $N >> M$. According to (\ref{tput1}), throughput achieved by SU $i$ will reach its maximum
value of 1 if its number of assigned channels is sufficiently large. Hence, we have proved the proposition.
\end{proof}

In practice, we do not need a very large number of channels to achieve the close-to-maximum throughput. In particular,
if each channel is available for secondary spectrum access with probability at least 0.8 then the throughput achieved
by a SU assigned three channels is not smaller than $1-(1-0.8)^3=0.992$, which is less than $1\%$ below the maximum
throughput. Note that after running Algorithm 1, we can establish the set of channels allocated to each SU, 
from which we calculate its throughput by using (\ref{tput1}). Then,
the total throughput of the secondary network can be calculated by summing the throughputs of all SUs. 
 When the number of channel is not sufficiently large, we can potentially improve the system throughput by allowing
overlapping channel assignment. We develop such an overlapping channel assignment algorithm in the next section. 

\section{Overlapping Channel Assignment}
\label{over}

Overlapping channel assignment can improve the network throughput by exploiting the multiuser diversity gain.
In particular, a channel assigned to only one SU cannot be exploited if it is being used by a nearby
primary user. However, if a particular channel is assigned to several SUs then it is more likely
that it can be exploited by at least one SU. However, when several
SUs attempt to access the same assigned channel, a MAC protocol is needed to resolve the access contention.
This MAC protocol incurs overhead that offsets the throughput gain due to the multiuser diversity. Hence, a sophisticated
channel assignment algorithm is needed to balance the protocol overhead and throughput gain.

\subsection{MAC Protocol}

\begin{figure}[!t]
\centering
\includegraphics[width=90mm]{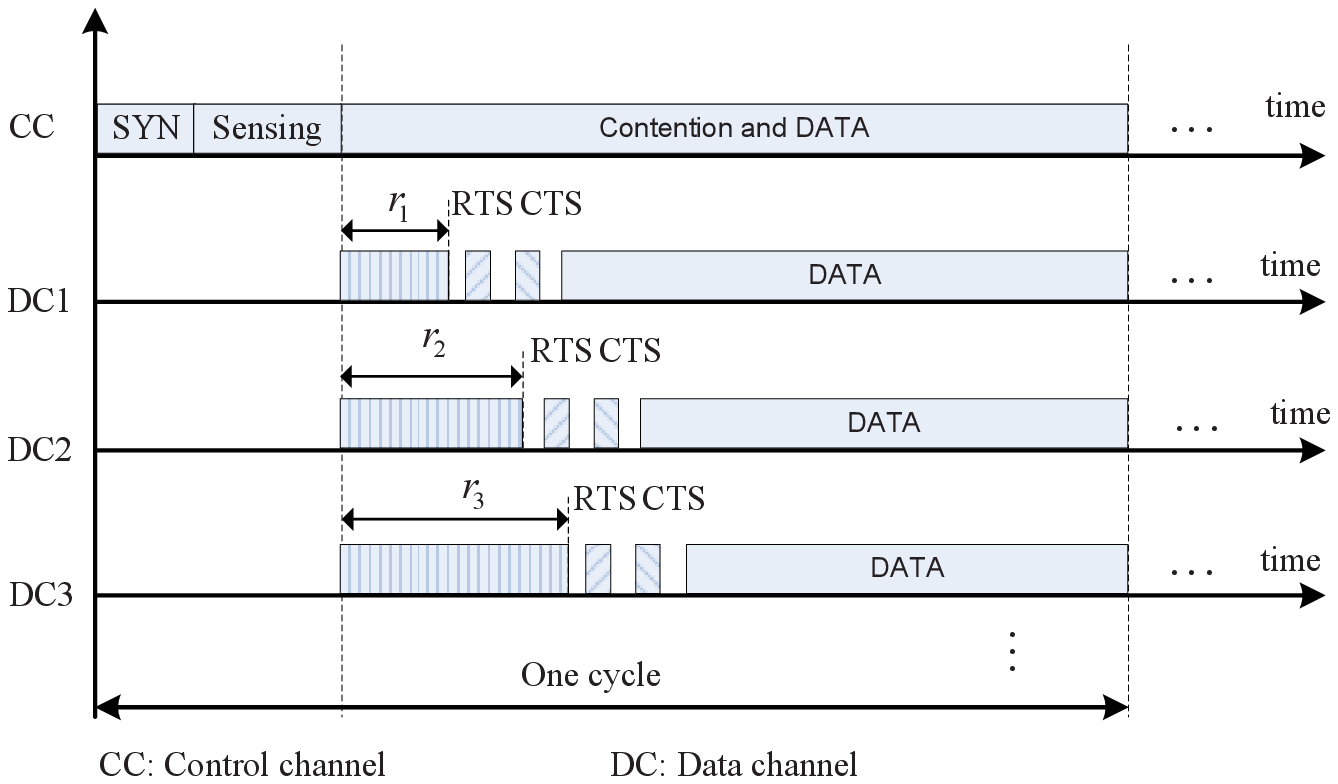}
\caption{Timing diagram for the proposed multi-channel MAC protocol.}
\label{MACoperation}
\end{figure}

Let $\mathcal{S}_i$ be the set of
channels  solely assigned  for SU $i$ and $\mathcal{S}_{i}^{\sf com}$ be the set of  channels assigned for SU $i$ and some other SUs. Let denote $\mathcal{S}_i^{\sf tot} = \mathcal{S}_i \cup \mathcal{S}_i^{\sf com}$, which is the set of all channels assigned to SU $i$. 
Assume that there is one control channel, which is always available and used for access contention resolution.
We consider the following MAC protocol run by any particular SU $i$, which belongs the class of synchronized MAC
protocol \cite{R2}. The MAC protocol is illustrated in Fig.~\ref{MACoperation} where
synchronization and sensing phases are employed before the channel contention and transmission phase in each cycle. 
A synchronization message is exchanged among SUs during the synchronization phase to establish
the same starting epoch of each cycle. After sensing the assigned channels in the sensing phase,
each SU $i$ proceeds as follows. If there is at least one channel in $\mathcal{S}_i$ available, then SU $i$ chooses
one of these available channels randomly for communication. If this is not the case, SU $i$ will choose one available channel
 in $\mathcal{S}_i^{\sf com}$ randomly (if there is any channel in this set available). For brevity, we simply call \textit{users} instead of
 \textit{SUs}  when there is no confusion. Then, it chooses
a random backoff value which is uniformly distributed in the interval $[0, W-1]$ (i.e.,  $W$ is the contention window) and 
starts decreasing its backoff counter while listening on the control channel.

\begin{table}
\centering
\caption{Channel Assignment Example (M=3, N =6) }
\label{table}
\begin{tabular}{|c|c|c|c|c|c|c|}
 \hline 
 & $\mathcal{S}_{1}$ & $\mathcal{S}_{2}$ & $\mathcal{S}_{3}$ & $\mathcal{S}_{1}^{\text{com}}$ & $\mathcal{S}_{2}^{\text{com}}$ & $\mathcal{S}_{3}^{\text{com}}$  \\
\hline \hline
C1 & x &  &   &  &  &  \\
\hline 
C2 &  & x &   &  &  &   \\
\hline 
C3 &  &  & x  &  &  &   \\
\hline 
C4 &  &  &   & x &  x &  \\
\hline 
C5 &  &  &  &  &  x  & x \\
\hline 
C6 &  &  &  &  x &  x  & x \\
\hline 
\end{tabular}
\end{table}

If it overhears transmissions of RTS/CTS from any other users, it will freeze from decreasing its backoff counter until the control
channel is free again. As soon as a user's backoff counter reaches zero, its transmitter transmits an RTS message
containing a chosen channel to its receiver. If the receiver successfully receives the RTS, it will reply with
CTS and user $i$ starts its communication on the chosen channel for the remaining of the cycle.
If the RTS/CTS message exchange fails due to collisions, the corresponding user will quit the contention and wait until the next cycle.
In addition, by overhearing RTS/CTS messages of neighboring users, which convey information about the channels chosen for communications, other users compared these channels with their chosen ones. 

Any user who has its chosen channel 
coincides with the overheard channels quits the contention and waits until the next cycle. Otherwise, it will continue to decrease its 
backoff counter before exchanging RTS/CTS messages. 
Note that the fundamental aspect that makes this MAC protocol different from
that proposed in \cite{Le11} is that in \cite{Le11} we assumed each winning user can use all available channels for
communications while at most one available channel can be exploited by hardware-constrained secondary users in the current paper.
Therefore, the channel assignment problem does not exist for the setting considered in  \cite{Le11}.

An example of overlapping channel assignment for three users and six channels is illustrated in Table I. 
Here, channel assignments are indicated by ``x'' in this table. As can be seen, each set $S_i$ contains
distinct channels while channels assigned to sets $S_i^{\text{com}}$ may be shared by more than one
users. For example, channel 5 (denoted as C5 in Table I) is shared by user 2 and user 3 and channel 6 is
shared by all three users in this assignment example.

\noindent
\textbf{Remark 1:} \textbf{We focus on the saturation-buffer scenario in this paper. In practice, cognitive radios may experience
dynamic packet arrivals which may result in empty buffers some time. Therefore, it is natural to allow
only backlogged users to perform sensing and access available channels in any cycle. In addition,
an efficient MAC protocol must allow users to fully utilize available channels during the data
transmission phase. Since the data transmission phase is quite large compared to a typical packet size,
we allow users to transmit several packets to completely fill the transmission phase in our MAC design. This
can be realized by requiring that only users who have sufficient data at the beginning of any particular
cycle can participate in the sensing and access contention processes. An alternative design would be to allow any backlogged
secondary users to participate in this process even if they do not have sufficient data to completely fill the data
transmission phase. These under-backlogged users can inform their neighbors by setting the network
 allocation vectors (NAVs) accordingly.  
Then we can allow other users to perform contention on the control channel during the 
 data transmission phase to utilize under-utilized channels, which are detected by decoding the NAVs
 of their neighbors. }

\subsection{Overlapping Channel Assignment Algorithm}

We develop an overlapping channel assignment algorithm that possesses two phases as follows.
We run Algorithm 1 to obtain the non-overlapping channel
assignment solution in the first phase. Then, we perform overlapping channel assignment
by allocating channels that have been assigned to some users
to other users in the second phase. The MAC protocol overhead typically increases when a larger number
of  users compete for the same channel. Therefore, to achieve
the optimal tradeoff between overhead and the multiuser diversity gain,
only small number of users should share any channel.

We devise a greedy overlapping channel assignment algorithm using the increase-of-throughput metric similar to that
employed in Algorithm 1. 
However, calculation of this metric exactly turns out to be a complicated
task. Hence, we employ an estimate of the increase-of-throughput, which is derived in the following
to perform channel assignment assuming that the MAC protocol overhead  is $\delta<1$. 
In fact, $\delta$ depends on the outcome of the channel assignment algorithm (i.e., sets of
channels assigned to different users). We will show
how to calculate $\delta$ and integrate it into this channel assignment algorithm later. 

Consider a case where channel $j$ is the common channel of users $i_1, i_2, \ldots, i_{\mathcal{MS}}$. Here,
$\mathcal{MS}$ is the number of users sharing this channel. We are interested in estimating
 the increase in throughput for a particular user $i$ if channel $j$ is assigned to this user.
 Indeed, this increase of throughput can be achieved because user $i$ may be able to exploit
 channel $j$ if this channel is not available or not used by other users $i_1, i_2, \ldots, i_{\mathcal{MS}}$.
 To estimate the increase of throughput, in the remaining of this paper we are only interested in
 a practical scenario where all $p_{ij}$ are close to 1 (e.g., at least 0.8). This would be a reasonable
 assumption given several recent measurements reveal that spectrum utilization of useful frequency bands
 is very low (e.g., less that $15\%$). Under this assumption,  
 we will show that the increase-of-throughput for user $i$ can be estimated as
\beqn \label{upith}
 \Delta T_{i}^{\mathcal{MS}, \text{est}} (j) =  
(1-1/\mathcal{MS})(1-\delta) p_{ij} \left( \prod_{h \in \mathcal{S}_i} \overline{p}_{ih} \right) \label{Del1} 
  \left(1-\prod_{h \in \mathcal{S}_i^{\text{com}}}\overline{p}_{ih} \right)  \sum_{k=1}^{\mathcal{MS}} \left[ \overline{p}_{i_k j} \left(\prod_{q=1, q \neq k}^{\mathcal{MS}}p_{i_q j}  \right) \right]  \\
+(1-\delta) p_{ij}  \prod_{h \in \mathcal{S}_i} \overline{p}_{ih}    \prod_{h \in \mathcal{S}_i^{\text{com}}}\overline{p}_{ih}  \label{Del2} 
 \prod_{q=1}^{\mathcal{MS}}p_{i_q j}   \prod_{q=1}^{\mathcal{MS}}\left(1-\prod_{h \in \mathcal{S}_{i_q}} \overline{p}_{i_q h}\right)   \hspace{0cm}\\
\hspace{0cm} + (1-1/\mathcal{MS})(1-\delta) p_{ij} \prod_{h \in \mathcal{S}_i} \overline{p}_{ih}  \left(1-\prod_{h \in \mathcal{S}_i^{\text{com}}}\overline{p}_{ih} \right)   
  \prod_{q=1}^{\mathcal{MS}}p_{i_q j}  \prod_{q=1}^{\mathcal{MS}}\left(1-\prod_{h \in \mathcal{S}_{i_q}} \overline{p}_{i_q h}\right). \label{Del3}
\eeqn

This estimation is obtained by listing all possible scenarios/events in which user $i$ can exploit channel $j$ to
increase its throughput. Because the user throughput is bounded by 1, we only count events that occur with
non-negligible probabilities. In particular, under the assumption that $p_{ij}$ are high (or $\overline{p}_{ij}$ are small)
we only count events whose probabilities have at most two such elements $\overline{p}_{ij}$ in the product. 
In addition, we can determine the increase of throughput for user $i$ by comparing its achievable throughput
before and after channel $j$ is assigned to it. It can be verified we have the following 
events for which the average increases of throughput are significant.
 
\begin{itemize}

\item Channel $j$ is available for all users $i$ and $i_q$, $q = 1, 2, \ldots, \mathcal{MS}$  except $i_k$ where $k = 1, 2, \ldots, \mathcal{MS}$. 
In addition, all channels in $S_i$ are not available and there is at least one channel in $\mathcal{S}_i^{\text{com}}$ available for user $i$. User $i$ can achieve a
maximum average throughput of $1-\delta$  by exploiting channel $j$, while its minimum average throughput before being assigned channel $i$
is at least $(1-\delta)/\mathcal{MS}$ (when user $i$ needs to share the available channel in $\mathcal{S}_i^{\text{com}}$ with $\mathcal{MS}$ 
other users). The increase of throughput for this case is at most $(1-1/\mathcal{MS})(1-\delta)$ and the upper-bound for the increase of
throughput of user $i$ is written in (\ref{Del1}).

\item Channel $j$ is available for user $i$ and all users $i_q$, $q = 1, 2, \ldots, \mathcal{MS}$ but each user $i_q$ uses other available channel in $\mathcal{S}_{i_q}$ for his/her transmission. Moreover, there is no channel in $\mathcal{S}_i^{\text{tot}}$ available. In this case, the increase of throughput for user $i$ is $1-\delta$ and the average increase of throughput of user $i$ is written in (\ref{Del2}).

\item Channel $j$ is available for user $i$ and all users $i_q$, $q = 1, 2, \ldots, \mathcal{MS}$ but each user $i_q$ uses other available channel in $\mathcal{S}_{i_q}$ for his/her transmission. Moreover, there is at least one channel in $\mathcal{S}_i^{\text{com}}$ available. In this case, the increase of throughput for user $i$ is upper-bounded by $(1-1/\mathcal{MS})(1-\delta)$ and the average increase of
throughput of user $i$ is written in (\ref{Del3}).

\end{itemize}

Detailed description of the algorithm is given in Algorithm 2. This algorithm has outer and inter loops where the outer loop
increases the parameter $h$, which represents the maximum of users allowed to share any particular channel (i.e., $\mathcal{MS}$ in
the above estimation of the increase of throughput) and the inner loop performs channel allocation for one particular value of $h=\mathcal{MS}$. 
In each assignment iteration of the inner loop, we assign one ``best'' channel $j$ to user $i$ that achieves maximum $\Delta T_{i}^{h,\text{est}} (j)$.
This assignment continues until the maximum $\Delta T_{i}^{h,\text{est}} (j)$ is less than a pre-determined number $\epsilon>0$.
As will be clear in the throughput analysis developed later, it is beneficial to maintain at least one channel in each set $S_i$.
This is because the throughput contributed by channels in $S_i$ constitutes a significant fraction of the total throughput.
Therefore, we will maintain this constraint when running Algorithm 2.

\subsection{Calculation of Contention Window }
\label{ConWinCal}

We show how to calculate contention window $W$ so that collision probabilities among contending secondary users are sufficiently small.
In fact, there is a trade-off between collision probabilities and the average overhead of the MAC protocol, which depends on $W$. In particular, larger values of $W$ reduce collision probabilities at the cost of higher protocol overhead and vice versa. Because there can be several collisions
during the contention phase each of which occurs if two or more users randomly choose the same value of backoff time.
In addition, the probability of the first collision is largest because the number of contending users decreases
for successive potential collisions.

Let $\mathcal{P}_c$ be the probability of the first collision. In the following, we determine contention window $W$ by imposing a
 constraint $\mathcal{P}_c \leq \epsilon_P $ where $\epsilon_P $ controls the collision probability and overhead tradeoff. Let us calculate $\mathcal{P}_c$ as a function of $W$ assuming that there are $m$ secondary 
 users in the contention phase. Without loss of generality, assume that the random backoff times of $m$  users are ordered as $r_1 \leq r_2 \leq \ldots \leq r_m$. The conditional probability of the first collision if there are $m$ users in the contention stage can be written as
\beqn
\label{Pfirstc}
\mathcal{P}_c^{(m)} &=& \sum _{j=2}^{m} \Pr \left( j \: \text{users collide} \right) \nonumber\\
&=& \sum_{j=2}^m \sum_{i=0}^{W-2} C_m^j \left( \frac{1}{W}\right)^j \left( \frac{W-i-1}{W}\right)^{m-j}
\eeqn
where each term in the double-sum represents the probability that $j$ users collide when they
choose the same backoff value equal to $i$. 
Hence, the probability of the first collision can be calculated as 
\beqn \label{pc}
\mathcal{P}_c = \sum_{m=2}^M \mathcal{P}_c^{(m)} \times \Pr\left\{m \: \text{ users contend}\right\},
\eeqn
where $\mathcal{P}_c^{(m)}$ is given in (\ref{Pfirstc}) and $\Pr\left\{m  \: \text{users contend}\right\}$
is the probability that $m$ users join the contention phase. To compute $\mathcal{P}_c$, we now
derive $\Pr\left\{m  \: \text{users contend}\right\}$. It can be verified that 
user $i$ joins contention if all channels in $\mathcal{S}_i$ are busy and there is at least one channel in $\mathcal{S}_i^{\sf com}$ available. The probability of this event can be written as
\beqn
\mathcal{P}_{\sf con}^{(i)} &=& \Pr \left\{ \text{all channels in} \: \mathcal{S}_i \: \text{ are busy}, 
 \exists ! \: \text{some channels in} \:\mathcal{S}_i^{\sf com} \: \text{are available }\right\} \nonumber\\
&=&\left( \prod_{j \in \mathcal{S}_i} \overline{p}_{ij} \right) \left( 1- \prod_{j \in \mathcal{S}_i^{\sf com}} \overline{p}_{ij} \right).
\eeqn
The probability of the event that $m $ users join the contention phase is 
\beqn \label{Pmusercon}
\Pr \left\{ m \: \text{users contend} \right\} = \sum_{n=1}^{C_M^m} \left( \prod_{i \in {\Lambda}_n} \mathcal{P}_{\sf con}^{(i)}\right) 
 \left( \prod_{j \in {\Lambda}_M \backslash {\Lambda}_n} \mathcal{\overline{P}}_{\sf con}^{(j)}\right) 
\eeqn
where ${\Lambda}_n$ is one particular set of $m$ users, ${\Lambda}_M$ is the set of all $M$ users ($\left\{ 1,2, \ldots , M \right\}$).
Substitute the result in (\ref{Pmusercon}) into (\ref{pc}), we can calculate $\mathcal{P}_c$. Finally,
we can determine $W$ as 
\beqn \label{Window}
W=  \min \left\{{W} \: \text{such that} \:  \mathcal{P}_c(W) \leq \epsilon_P \right\}
\eeqn
where for clarity we denote $\mathcal{P}_c(W)$, which is given in (\ref{pc}) as a function of $W$.

\subsection{Calculation of MAC Protocol Overhead}
\label{Overcal}

Let $r$ be the average value of the backoff value chosen by any SU.
Then, we have $r = (W-1)/2$ because the backoff counter value is uniformly chosen
in the interval $[0,W-1]$. As a result, average overhead can be calculated as follows:
\beqn \label{overhead}
\delta\left(W\right) = \frac { \left[ W-1 \right]\theta/2 + t_{\sf RTS} + t_{\sf CTS} + 3 t_{\sf SIFS} + t_{\sf SEN}+ t_{\sf SYN}} {\sf T_{\sf cycle}}
\eeqn
where $\theta$ is the time corresponding to one backoff unit;  $t_{\sf RTS}$,  $t_{\sf CTS}$, $t_{\sf SIFS}$
are the corresponding time of RTS, CTS and SIFS (i.e., short inter-frame space) messages; $t_{\sf SEN}$ is the
sensing time; $t_{\sf SYN}$ is the transmission time of the synchronization message; 
and $\sf T_{\sf cycle}$ is the cycle time.

\subsection{Update $\delta$ inside Algorithm 2}

Because the overhead $\delta$ depends on the channel assignment outcome, which is not known when we are 
running Algorithm 2. Therefore, in each allocation step we update $\delta$ based on the current channel assignment outcome.
Because $\delta$ does not change much in two consecutive allocation decisions, Algorithm 2 runs smoothly in practice.

\subsection{Practical Implementation Issues}

\textbf{To perform channel assignment, we need to know $p_{ij}$ for all users and channels. Fortunately, we only need to perform estimation 
of $p_{ij}$ once these values change,
which would be infrequent in practice. These estimation and channel assignment tasks can be performed by one secondary node or collaboratively
performed by several of them. For example, for the secondary network supporting communications between $M$ secondary users and a single secondary
BS, the BS can take the responsibility of estimating $p_{ij}$ and performing channel assignment. Once the channel assignment solution
has been determined and forwarded to all secondary users, each secondary user will perform spectrum sensing and run the underlying MAC
protocol to access the spectrum in each cycle. }

\textbf{It is emphasized again that while sensing and MAC protocol are performed and run in every cycle, 
estimating of $p_{ij}$ and performing channel assignment (given these $p_{ij}$) are only performed if the values of $p_{ij}$ change,
which should be infrequent. Therefore, it would be affordable to estimate $p_{ij}$ accurately by employing sufficiently long
sensing time. This is because for most spectrum sensing schemes including an energy detection scheme, mis-detection and false alarm probabilities
tend to zero when sensing time increases for a given sampling frequency \cite{Yu09}, \cite{R1}.}

\section{Performance Analysis}
\label{Tputderive}

Suppose we have run Algorithm 2 and obtained the set of users $U_j$ associated with each allocated channel $j$.
From this, we have the corresponding sets $\mathcal{S}_i$ and $\mathcal{S}_i^{\text{com}}$ for each user $i$.
Given this channel assignment outcome, we derive the throughput in the following assuming that there is no
collision due to MAC protocol access contention. We will show that by appropriately choosing
contention parameters for the MAC protocol, the throughput analysis under this assumption achieves accurate 
results. 

\subsection{Throughput Analysis}
\label{tputana}

Because the total throughput is the sum of throughput of all users, it is sufficient to analyze the throughput of one
particular user $i$. We will perform the throughput analysis by considering all possible sensing outcomes performed by
the considered user $i$ for its assigned channels. 
We will have the following cases, which correspond to different achievable throughput for the considered user.
\begin{itemize}
\item{ Case 1: If there is at least one channel in $\mathcal{S}_i$ available, then user $i$ will exploit this available
channel and achieve the throughput of one. Here, we have
\beqn
T_i \left\{\text{Case 1} \right\}  = \Pr\left\{\text{Case 1} \right\} = 1-\prod\limits_{j\in \mathcal{S}_i} \bar{p}_{ij}.
\eeqn }
\item{\text{Case 2}: If no channel in $\mathcal{S}_i^{\sf tot} $ is available for user $i$, then the achievable throughput
of user $i$ is zero. This scenario occurs with following probability
\beqn
\Pr\left\{\text{Case 2}\right\} = \prod \limits _{j\in \mathcal{S}_i^{\sf tot}} \bar{p}_{ij}.
\eeqn}

\item{$\text{Case 3}$: In this case, we consider scenarios where all channels in $\mathcal{S}_i$ are not 
available; there is at least one channel in $\mathcal{S}_i^{\text{ com}}$ available, and user $i$ chooses the available channel $j$ for transmission.
Suppose that channel $j$ is shared  by $\mathcal{MS}_j$ secondary users including user $i$ (i.e., $\mathcal{MS}_j = |U_j|$). 
There are four possible groups of users $i_k$, $k=1, \ldots, \mathcal{MS}_j$ sharing channel $j$, which are described in
the following
\begin{itemize}
\item{ \textbf{Group I}: channel $j$ is available for user $i_k$ and user $i_k$ has at least 1 channel in $\mathcal{S}_{i_k}$ available.}
\item{ \textbf{Group II}: channel $j$ is not available for user $i_k$. }
\item{ \textbf{Group III}: channel $j$ is available for user $i_k$, all channels in $\mathcal{S}_{i_k}$ are not available and there is another channel
 $j'$ in $\mathcal{S}_{i_k}^{\text{com}}$ available for user $i_k$. In addition, user $i_k$ chooses channel $j'$ for transmission in the contention stage.}
\item{ \textbf{Group IV}: channel $j$ is available for user $i_k$, all channels in $\mathcal{S}_{i_k}$ are not available. In addition, user $i_k$  chooses channel $j$ for transmission in the contention stage. Hence, user $i_k$ competes with user $i$ for channel $j$.}

\end{itemize}

The throughput achieved by user $i$ in this case can be written as
\beqn
\label{Tputa1delta}
T_i\left( \text{ Case 3} \right) = 
 (1-\delta) \Theta_i \sum \limits_{A_1 = 0}^{\mathcal{MS}_j} \sum \limits_{A_2 = 0}^{\mathcal{MS}_j-A_1} \sum \limits_{A_3=0}^{\mathcal{MS}_j-A_1-A_2} 
 \Phi_1(A_1) \Phi_2(A_2)  \Phi_3(A_3) \Phi_4(A_4)
\eeqn
where 
\begin{itemize}
\item $\Theta_i$ is the probability that all channels in $\mathcal{S}_i$ are not available and user $i$ chooses 
some available channel $j$ in $\mathcal{S}_i^{\text{com}}$ for transmission.
\item $\Phi_1(A_1)$ denotes the probability that there are $A_1$ users belonging to Group I described above
among $\mathcal{MS}_j$ users sharing channel $j$.
\item $\Phi_2(A_2)$ represents the probability that there are $A_2$ users belonging to Group II 
among $\mathcal{MS}_j$ users sharing channel $j$.
\item $\Phi_3(A_3)$ describes the probability that there are $A_3$ users belonging to Group III
among $\mathcal{MS}_j$ users sharing channel $j$.
\item $\Phi_4(A_4)$ denotes the probability that there are $A_4 = \mathcal{MS}_j-A_1-A_2-A_3$ 
remaining users belonging to Group IV  scaled by $1/(1+A_4)$ where $A_4$ is the number of users 
excluding user $i$ competing with user $i$ for channel $j$.
\end{itemize}
We now proceed to calculate these quantities. We have
\beqn \label{group0}
\Theta_i =  \prod_{k \in \mathcal{S}_i} \overline{p}_{ik} \sum \limits _{B_i =1}^{H_i} \sum \limits_{h=1}^{C_{H_i}^{B_i}} \sum \limits_{j \in {\Psi}^h_i}  \frac{1}{B_i} \prod_{j_1 \in \Psi^h_i } p_{ij_1} \!\!\!\!\!\!\!\!\!\prod_{j_2 \in \mathcal{S}_i^{\text{com}} \backslash \Psi^h_i} \!\!\!\!\!\!\!\!\overline{p}_{ij_2} 
\eeqn
where $H_i$ denotes the number of channels in $\mathcal{S}_i^{\text{com}}$. The first product term in (\ref{group0}) represents
the probability that all channels in $\mathcal{S}_i$ are not available for user $i$. The second term in (\ref{group0}) describes the
probability that user $i$ chooses an available channel $j$ among $B_i$ available channels in $\mathcal{S}_i^{\text{com}}$ for transmission.
Here, we consider all possible subsets of $B_i$ available channels and for one such particular case $\Psi^h_i$ describes the corresponding
set of $B_i$ available channels.
\beqn
\Phi_1(A_1) = \sum \limits_{c_1=1}^{C_{\mathcal{MS}_j}^{A_1}} \prod \limits_{m_1 \in  \Omega_{c_1}^{(1)}  } \left(p_{m_1j} \left(1-\prod \limits_{l \in \mathcal{S}_{m_1}}\overline{p}_{m_1 l}\right)\right).   \label{groupI}
\eeqn
In (\ref{groupI}), we consider all possible subsets of size $A_1$ belonging to Group I (there are $C_{\mathcal{MS}_j}^{A_1}$ such subsets).
Each term inside the sum represents the probability for the corresponding event whose set of $A_1$ users is denoted by $\Omega_{c_1}^{(1)}$.
\beqn
\Phi_2(A_2) = \sum \limits_{c_2=1}^{C_{\mathcal{MS}_j-A_1}^{A_2}} \prod \limits_{m_2 \in \Omega_{c_2}^{(2)}  }  \overline{p}_{m_2j}. \label{groupII} 
\eeqn
In (\ref{groupII}), we capture the probability that channel $j$ is not available for $A_2$ users in group II whose possible sets are denoted by $\Omega_{c_2}^{(2)}$. 
\beqn
\Phi_3(A_3) = \sum \limits_{c_3=1}^{C_{\mathcal{MS}_j-A_1-A_2}^{A_3}} \prod \limits_{m_3 \in \Omega_{c_3}^{(3)}  } \left(p_{m_3j} \prod_{l_3 \in \mathcal{S}_{m_3}} \overline{p}_{m_3l_3}\right)  \label{1groupIII} \hspace{3.0cm} \\ 
\hspace{0cm} \times   \left[  \sum \limits_{n=0}^{\beta} \sum_{q=1}^{C_{\beta}^n} 
\prod_{h_1 \in \mathcal{S}^{\text{com},q}_{j,m_3} }  p_{m_3 h_1} 
\prod_{h_2 \in \overline{\mathcal{S}}^{\text{com},q}_{j,m_3} }  \overline{p}_{m_3 h_2}  
  \left( 1- \frac{1}{ n+1} \right) \right]. \label{2groupIII} 
\eeqn
For each term in (\ref{1groupIII}) we consider different possible subsets of $A_3$ users, which are denoted by $\Omega_{c_3}^{(3)}$.
Then, each term in (\ref{1groupIII})  represents the probability that channel $j$ is available for each user $m_3 \in \Omega_{c_3}^{(3)}$ 
while all channels in  $\mathcal{S}_{m_3}$ for the user $m_3$ are not available. 
In (\ref{2groupIII}), we consider all possible sensing outcomes for channels in $\mathcal{S}_{m_3}^{\text{com}}$ performed by user
 $m_3 \in \Omega_{c_3}^{(3)}$.
In addition, let $\mathcal{S}^{\text{com}}_{j,m_3} = \mathcal{S}_{m_3}^{\text{com}} \backslash \left\{ j \right\}$  
and $\beta = |\mathcal{S}^{\text{com}}_{j,m_3}|$.
Then, in (\ref{2groupIII}) we consider all possible scenarios in which there are $n$ channels in $\mathcal{S}^{\text{com}}_{j,m_3}$ available; and user $m_3$ chooses a channel different from channel $j$ for transmission (with probability $\left( 1- \frac{1}{ n+1} \right)$) where
$\mathcal{S}^{\text{com}}_{j,m_3} = \mathcal{S}^{\text{com},q}_{j,m_3} \cup \overline{\mathcal{S}}^{\text{com},q}_{j,m_3} $ and
$\mathcal{S}^{\text{com},q}_{j,m_3} \cap \overline{\mathcal{S}}^{\text{com},q}_{j,m_3} = \emptyset$.
\beqn
\Phi_4(A_4) =  \left(\frac{1}{1+ A_4}\right) \prod \limits_{m_4 \in \Omega^{(4)} } \left(p_{m_4j} \prod_{l_4 \in \mathcal{S}_{m_4}} \overline{p}_{m_4l_4}\right)  \label{1groupIV}   \hspace{3cm}\\
\times \left[ \sum \limits_{m=0}^{\gamma} \sum_{q=1}^{C_{\gamma}^{m}}   
\prod_{h_1 \in \mathcal{S}^{\text{com},q}_{j,m_4} }  p_{m_4 h_1} 
\prod_{h_2 \in \overline{\mathcal{S}}^{\text{com},q}_{j,m_4} }  \overline{p}_{m_4 h_2}  
\left( \frac{1}{ m+1} \right)  \right].
 \label{2groupIV} 
\eeqn
The sensing outcomes captured in (\ref{1groupIV}) and (\ref{2groupIV}) are similar to those in (\ref{1groupIII}) and (\ref{2groupIII}).
However, given three sets of $A_1$, $A_2$, and $A_3$ users, the set $\Omega^{(4)}$ can be determined whose size is $|\Omega^{(4)}| = A_4$. 
Here, $\gamma$ denotes cardinality of the set $\mathcal{S}_{j,m_4}^{\text{com}} = \mathcal{S}_{m_4}^{\text{com}} \backslash \left\{ j \right\}$.
Other sets are similar to those in (\ref{1groupIII}) and (\ref{2groupIII}). However, all users in $\Omega^{(4)}$ choose channel $j$ for transmission
in this case. Therefore, user $i$ wins the contention with probability $1/(1+A_4)$ and its achievable throughput is $(1-\delta)/(1+A_4)$.

}
\end{itemize}

Summarizing all considered cases, the throughput achieved by user $i$ is given as
\beqn
T_i = T_i \left\{\text{Case 1} \right\} + T_i \left\{ \text{Case 3} \right\} .
\eeqn
In addition, the total throughput of the secondary network $\mathcal{T}$ is the sum of throughputs achieved by all secondary users.

\subsection{Impacts of Contention Collision}
\label{conovh}

We have presented the saturation throughput analysis assuming that there is no contention collision.
Intuitively, if the MAC protocol is designed such that collision probability is sufficiently small
then the impact of collision on the throughput performance would be negligible. For our MAC protocol,
users  perform contention resolution in Case 3 considered in the previous throughput analysis,
which occurs with a small probability. Therefore, if the contention window in (\ref{Window}) is chosen
for a sufficiently small $\epsilon_P$, then contention collisions would have negligible impacts on the network throughput.
We state this intuitive result formally in the following proposition.

\vspace{0.1cm}
\noindent
\textbf{Proposition 2:} The throughput $\mathcal{T}$ derived in the previous sub-section has an error, which can be upper-bounded as 
\beq \label{error}
E_t \leq  \epsilon_P \sum_{i=1}^M \prod\limits_{j\in \mathcal{S}_i} \bar{p}_{ij} \left( 1 - \prod\limits_{j\in 
\mathcal{S}_i^{\text{com}}} \bar{p}_{ij}  \right)
\eeq
where $\epsilon_P$ is the target collision probability used to determine the contention window in (\ref{Window}).

\begin{proof}
As discussed above, contention collision can only occur in Case 3 of the previous throughput analysis. The probability covering all
possible events for user $i$ in this case is $\prod\limits_{j\in \mathcal{S}_i} \bar{p}_{ij} \left( 1 - \prod\limits_{j\in 
\mathcal{S}_i^{\text{com}}} \bar{p}_{ij}  \right)$. In addition, the maximum average throughput that a particular user $i$ 
can achieve is $1-\delta<1$ (as no other users contend with user $i$ to exploit a chosen channel). In addition, if contention collision
happens then user $i$ will quit the contention and may experience a maximum average throughput loss of $1-\delta$ compared to the 
ideal case with no contention collision. Also, collision probabilities of all potential collisions is bounded above by $\epsilon_P$.
Therefore, the average error due to the proposed throughput analysis
can be upper-bounded as in (\ref{error}).
\end{proof}

To illustrate the throughput error bound presented in this proposition, let us consider an example where
$\bar{p}_{ij} \leq 0.2$ and $\epsilon_P \leq 0.03$. Because the sets $S_i$ returned by Algorithm 2 contain at least
one channel, the throughput error can be bounded by $M \times 0.2 \times 0.03 = 0.006M$. In addition, the total
throughput will be at least $\sum_{i=1}^M \left(1 - \prod\limits_{j\in \mathcal{S}_i} \bar{p}_{ij}\right) \geq 0.8M$ if
we only consider throughput contribution from Case 1.
Therefore, the relative throughput error can be upper-bounded by $0.006M/0.8M \approx 0.75 \%$, which is quite negligible. 
This example shows that the proposed throughput analytical model is very accurate in most practical settings.

\subsection{Complexity Analysis}
\label{Compx}

\textbf{
We analyze the complexity of Algorithm 1 and Algorithm 2 in this subsection.
Let us proceed by analyzing the steps taken in each iteration in Algorithm 1.
To determine the best assignment for the first channel, we have to search over $M$ SUs and $N$ channels, which involves $MN$ cases. Similarly, to assign the second channel, we need to perform searching over secondary users and $N-1$ channels (one channel is already assigned in
the first iteration). Hence, the second assignment involves $M\left(N-1\right)$ cases. Similar analysis can be applied for other assignments in later iterations. In summary, the total number of cases involved in assigning all channels to $M$ SUs is $M\left(N+\ldots+2+1\right) = MN\left(N+1\right)/2$,
which is $O(MN^2)$. In Algorithm 1, the increase of throughput used in the search is calculated by using (\ref{Tlem1}).}

\textbf{
In Algorithm 2, we run Algorithm 1 in the first phase then perform further overlapping channel assignments using Algorithm 2 in the second phase.
Hence, we need to analyze the complexity involved in the second phase (i.e., Algorithm 2).
In Algorithm 2, we increase the parameter $h$ from 1 to $M-1$ over iterations of the while loop to allow increasing number of users
to share one channel. For a particular value of $h$, we search over the channels that have been shared by $h$ users and over
all $M$ users. Therefore, we have $NM$ cases to consider for each value of $h$ each of which requires to calculate the corresponding increase
of throughput using (\ref{upith}). Therefore, the worst case complexity of the second phase is $NM(M-1)$, which is $O(NM^2)$.
Considering the complexity of both phases, the complexity of Algorithm 2 is $O(MN^2+NM^2)=O(MN(M+N))$,
which is much lower than that of the optimal brute-force search algorithm ($O(2^{NM})$).}


\section{Further Extensions and Design Issues}
\label{Poten}

\subsection{Fair Channel Assignment}
\label{Fairness}

We extend the channel assignment considering max-min fairness, which maximizes the minimum throughput achieved by all
secondary users. Note that max-min a popular fairness criterion that has been widely used in wireless resource allocation.
Throughput performance achieved under fairness criteria such as proportional fairness will be in between those under throughput
maximization and max-min fairness \cite{tass05}, which, therefore, provide useful performance bounds for other fair optimization objectives.
Toward this end, the max-min channel assignment problem can be stated as follows:
\vspace{0.0cm}
\beqn
\label{Tput_fair}
&& \mathop {\max_i } \mathop {\min }_\textbf{x} {T_i}.  \label{obj1} 
\eeqn
Intuitively, the max-min fairness criterion tends to allocate more radio resources for ``weak'' users to balance the throughput
performance among all users. Thanks to the exact throughput analytical model developed in Section VI.A, the optimal solution
of the optimization problem (\ref{obj1}) can be found by the exhaustive search. Specifically, we can enumerate all possible
channel allocations and calculate their corresponding throughput performance. Then, the optimal solution is the one that
achieves the largest value in (\ref{obj1}). As being discussed before this exhaustive search has extremely high computational complexity.

To resolve this complexity issue, we devise greedy fair non-overlapping and overlapping channel assignment algorithms, which are 
described in Algorithm 3 and Algorithm 4, respectively. In Section VIII, we compare the performance
of these algorithms with that of the optimal exhaustive search algorithm. These algorithms are different from
Algorithm 1 and Algorithm 2 mainly in the way we choose the user to allocate one ``best'' channel in each iteration.
In Algorithm 3, we find the set of 
users who achieve a minimum throughput in each iteration. For each user in this set, we find one available channel
that results in the highest increase of throughput. Then, we assign the ``best'' channel that achieves the maximum
increase of throughput considering all throughput-minimum users. Therefore, this assignment attempts to increase the throughput
of a weak user while exploiting the multiuser diversity.

In Algorithm 4, we first run Algorithm 3 to obtain non-overlapping sets of channels for all users. Then, we seek to improve
the minimum throughput by performing overlapping channel assignments. 
In particular, we find the minimum-throughput user and an overlapping channel assignment that results in the largest increase in its throughput.
The algorithm terminates when there is no such overlapping channel assignment. The search of an overlapping channel assignment
in each iteration of Algorithm 4 is performed in Algorithm 5. Specifically, we sequentially search over channels which have already been
allocated for a single user or shared by several users (i.e., channels in separate and common sets, respectively). Then, we update the
current temporary assignment with a better one (if any) during the search. This search requires throughput calculations for which
we use the analytical model developed in Subsection \ref{tputana} with the MAC protocol overhead, $\delta<1$ derived in \ref{Overcal}.
It can be observed that the proposed throughput analysis is very useful since it can be used to evaluate the performance of any
channel assignment solution and to perform channel assignments in greedy algorithms.

\subsection{Throughput Analysis under Imperfect Sensing}
\label{Imper_sens}

We extend the throughput analysis considering imperfect sensing in this subsection.
The same synchronized MAC protocol described in Section V.A is assumed here.
In addition, the MAC protocol overhead can be calculated as presented in Section~ \ref{Overcal} where the contention window $W$ 
is determined as described in Section \ref{ConWinCal}. 
There are two important performance measures,
which are used to quantify the sensing performance, namely detection and false alarm probabilities. 
Let $\mathcal{{P}}_d^{ij}$ and $\mathcal{{P}}_f^{ij}$ be detection and false alarm probabilities, respectively
of SU $i$ on channel $j$. In particular,
detection event occurs when a secondary link successfully senses a busy channel and false alarm
represents the situation when a spectrum sensor returns a busy state for an idle channel (i.e., a transmission opportunity
is overlooked). Also, let us define $\mathcal{{P}}_d^{ij} = 1-\mathcal{\overline{P}}_d^{ij}$ and $\mathcal{\overline{P}}_f^{ij} =  1- \mathcal{{P}}_f^{ij}$. Under imperfect sensing, there are four possible scenarios for channel $j$ and secondary user $i$.

\begin{itemize}
\item {Scenario I}: A spectrum sensor indicates that channel $j$ is available and the nearby PU is not using channel $j$ (i.e., correct
sensing). This scenario occurs with the probability $\mathcal{\overline{P}}_f^{ij}p_{ij}$.

\item {Scenario  II}: A spectrum sensor indicates that channel $j$ is available and the nearby PU is using channel $j$ (i.e., mis-detection). This scenario occurs with the probability $\mathcal{\overline{P}}_d^{ij} \overline{p}_{ij}$. In this case, potential transmission of secondary user $i$ will collide with that of the nearby primary user. We assume that both transmissions from SU $i$ and the nearby PU fail.

\item {Scenario  III}: A spectrum sensor indicates that channel $j$ is not available and the nearby PU is using channel $j$ (i.e., correct detection). 
This scenario occurs with the probability $\mathcal{P}_d^{ij} \overline{p}_{ij}$.

\item {Scenario IV}: A spectrum sensor indicates that channel $j$ is not available and the nearby PU is not using channel $j$ (i.e., false alarm). 
 This scenario occurs with the probability $\mathcal{P}_f^{ij} p_{ij}$ and the channel opportunity is overlooked.

\end{itemize}

Since SUs make channel access decisions based on their sensing outcomes, the first two scenarios can result in spectrum access on channel 
$j$ by SU $i$. Moreover, spectrum access in scenario one actually lead to successful data transmission. 
Let us define $\mathcal{P}_{\text{idle}}^{ij}=\mathcal{\overline{P}}_f^{ij}p_{ij}+\mathcal{\overline{P}}_d^{ij}\overline{p}_{ij}$ and $\mathcal{P}_{\text{busy}}^{ij}=1-\mathcal{P}_{\text{idle}}^{ij}$ as the probabilities under which SU $i$ may and may not access channel 
$j$, respectively. Since the total throughput is the sum of throughput of all users, it is sufficient to analyze the throughput of one
particular user $i$. To analyze the throughput of user $i$, we consider the following cases.
\begin{itemize}
\item{ Case 1: There is at least one channel in $\mathcal{S}_i$ available and user $i$ actually
chooses one of these available channels for its transmission. User $i$ can achieve throughput of one in such a successful access,
which occurs with the following probability
\beqn
T_i \left\{\text{Case 1} \right\}  = \Pr\left\{\text{Case 1} \right\} = \sum \limits_{k_1=1}^{\left| \mathcal{S}_i\right|} \sum \limits_{l_1=1}^{C_{\left| \mathcal{S}_i\right|}^{k_1}} 
\prod \limits_{j_1 \in \mathcal{S}_i^{l_1}} p_{ij_1} \prod \limits_{j_2 \in \mathcal{S}_i \backslash \mathcal{S}_i^{l_1}} \overline{p}_{ij_2} \label{Case1_1} \hspace{5.0cm} \\
\sum \limits_{k_2=1}^{k_1} \sum \limits_{l_2=1}^{C_{k_1}^{k_2}} 
\prod \limits_{j_3 \in \mathcal{S}_i^{l_2}} \mathcal{\overline{P}}_f^{ij_3} \prod \limits_{j_4 \in \mathcal{S}_i^{l_1} \backslash \mathcal{S}_i^{l_2}} \mathcal{P}_f^{ij_4} \label{Case1_2} \hspace{4.0cm}\\
\sum \limits_{k_3 = 0}^{\left|\mathcal{S}_i\right|-k_1} \sum \limits_{l_3=1}^{C_{\left|\mathcal{S}_i\right|-k_1}^{k_3}} \frac{k_2}{k_2+k_3} \prod \limits_{j_5 \in \mathcal{S}_i^{l_3}} \mathcal{\overline{P}}_d^{ij_5} \prod \limits_{j_6 \in \mathcal{S}_i \backslash \mathcal{S}_i^{l_1} \backslash \mathcal{S}_i^{l_3}} \mathcal{P}_d^{ij_6}. \label{Case1_3} \hspace{0.0cm}
\eeqn 

The quantity (\ref{Case1_1}) represents the
probability that there are $k_1$ actually available channels in $\mathcal{S}_i$ (which may or may not be correctly sensed by SU $i$). Here, $\mathcal{S}_i^{l_1}$ denotes a particular set of $k_1$ actually available channels whose index is $l_1$.
In addition, the quantity (\ref{Case1_2}) describes the
probability that there are $k_2$ available channels  actually indicated by sensing (the remaining available channels are overlooked due to
sensing errors) where $\mathcal{S}_i^{l_2}$ denotes the $l_2$-th set with $k_2$ available channels. For the quantity in (\ref{Case1_3}),
$k_3$ denotes the number of channels that are not actually available but the sensing outcomes indicate they are available (i.e., due to
mis-detection). Moreover, $k_2/(k_2+k_3)$ represents the probability that SU $i$ chooses the actually available channel for transmission
given its sensing outcomes indicate $k_2+k_3$ available channels. The remaining quantity in (\ref{Case1_3}) describes the probability
that the sensing outcomes due to SU $i$ incorrectly indicates $k_3$ available channels. }
\item{\text{Case 2}: There can be some channels in $\mathcal{S}_i^{\sf tot}$ available for user $i$ but
the sensing outcomes indicate that all channels are busy. As a result, user $i$ does not attempt to access any channel. This scenario occurs with following probability
\beqn
\Pr\left\{\text{Case 2}\right\} =  \sum \limits_{k_1=0}^{\left| \mathcal{S}_i^{\sf tot} \right|} \sum \limits_{l_1=1}^{C_{\left| \mathcal{S}_i^{\sf tot} \right|}^{k_1}} 
\prod \limits_{j_1 \in \mathcal{S}_i^{l_1}} \mathcal{P}_f^{ij_1} p_{ij_1} \prod \limits_{j_2 \in \mathcal{S}_i^{\sf tot} \backslash \mathcal{S}_i^{l_1}} \overline{p}_{ij_2}.
\eeqn}
The achievable throughput of user $i$ is zero in this case.

\item{$\text{Case 3}$: All channels in $\mathcal{S}_i$ are indicated as not available by sensing; there is at least one channel in $\mathcal{S}_i^{\text{ com}}$ indicated as available by sensing, and user $i$ chooses an actually available channel $j$ for transmission. 
Suppose that channel $j$ is shared  by $\mathcal{MS}_j$ secondary users including user $i$ (i.e., $\mathcal{MS}_j = |U_j|$). 
There are four possible groups of users $i_k$, $k=1, \ldots, \mathcal{MS}_j$ sharing channel $j$, which are described in
the following
\begin{itemize}
\item{ \textbf{Group I}: channel $j$ is available for user $i_k$ and user $i_k$ has at least 1 channel in $\mathcal{S}_{i_k}$ available as indicated by sensing.}
\item{ \textbf{Group II}: channel $j$ is indicated as not available for user $i_k$ by sensing. }
\item{ \textbf{Group III}: channel $j$ is available for user $i_k$, all channels in $\mathcal{S}_{i_k}$ are not available and there is another channel
 $j'$ in $\mathcal{S}_{i_k}^{\text{com}}$ available for user $i_k$ as indicated by sensing. In addition, user $i_k$ chooses channel $j'$ for transmission in the contention stage.}
\item{ \textbf{Group IV}: channel $j$ is available for user $i_k$, all channels in $\mathcal{S}_{i_k}$ are not available as indicated by sensing. In addition, user $i_k$  chooses channel $j$ for transmission in the contention stage. Hence, user $i_k$ competes with user $i$ for channel $j$.}

\end{itemize}

The throughput achieved by user $i$ in this case can be written as
\beqn
\label{Tputa1delta1}
T_i\left( \text{ Case 3} \right) = 
 (1-\delta) \Theta_i \sum \limits_{A_1 = 0}^{\mathcal{MS}_j} \sum \limits_{A_2 = 0}^{\mathcal{MS}_j-A_1} \sum \limits_{A_3=0}^{\mathcal{MS}_j-A_1-A_2} 
 \Phi_1(A_1) \Phi_2(A_2)  \Phi_3(A_3) \Phi_4(A_4).
\eeqn
Here, we use the same notation as for the perfect sensing where
\begin{itemize}
\item $\Theta_i$ is the probability that all channels in $\mathcal{S}_i$ are indicated as not available by sensing and user $i$ chooses 
some available channel $j$ in $\mathcal{S}_i^{\text{com}}$ as indicated by sensing for transmission.
\item $\Phi_1(A_1)$ denotes the probability that there are $A_1$ users belonging to Group I described above
among $\mathcal{MS}_j$ users sharing channel $j$.
\item $\Phi_2(A_2)$ represents the probability that there are $A_2$ users belonging to Group II 
among $\mathcal{MS}_j$ users sharing channel $j$.
\item $\Phi_3(A_3)$ describes the probability that there are $A_3$ users belonging to Group III
among $\mathcal{MS}_j$ users sharing channel $j$.
\item $\Phi_4(A_4)$ denotes the probability that there are $A_4 = \mathcal{MS}_j-A_1-A_2-A_3$ 
remaining users belonging to Group IV  scaled by $1/(1+A_4)$ where $A_4$ is the number of users 
excluding user $i$ competing with user $i$ for channel $j$.
\end{itemize}
We now proceed to calculate these quantities. We have
\beqn \label{group1_0}
\Theta_i =  \sum \limits_{k_1=0}^{\left| \mathcal{S}_i\right|} \sum \limits_{l_1=1}^{C_{\left| \mathcal{S}_i\right|}^{k_1}} 
\prod \limits_{j_1 \in \mathcal{S}_i^{l_1}} \mathcal{P}_f^{ij_1} p_{ij_1} \prod \limits_{j_2 \in \mathcal{S}_i \backslash \mathcal{S}_i^{l_1}} \overline{p}_{ij_2} \label{group1_00} \hspace{6.0cm} \\
\sum \limits _{k_2 =1}^{H_i} \sum \limits_{l_2=1}^{C_{H_i}^{k_2}} 
\prod_{j_3 \in \Psi^{l_2}_i } p_{ij_3} \prod_{j_4 \in \mathcal{S}_i^{\text{com}} \backslash \Psi^{l_2}_i} \overline{p}_{ij_4} \label{group1_01} \hspace{5.0cm} \\
\sum \limits_{k_3=1}^{k_2} \sum \limits_{l_3=1}^{C_{k_2}^{k_3}}  \sum \limits_{j \in \Gamma_1^k} 
\prod \limits_{j_5 \in \Gamma_1^{l_3}} \mathcal{\overline{P}}_f^{ij_5} \prod \limits_{j_6 \in \Psi^{l_2}_i \backslash \Gamma_i^{l_3}} \mathcal{P}_f^{ij_6} \label{group1_02} \hspace{4.0cm} \\
\sum \limits_{k_4 = 0}^{H_i-k_2} \sum \limits_{l_4=1}^{C_{H_i-k_2}^{k_4}} \frac{1}{k_3+k_4} \prod \limits_{j_7 \in \Gamma_2^{l_4}} \mathcal{\overline{P}}_d^{ij_7} \prod \limits_{j_8 \in \mathcal{S}_i^{\text{com}} \backslash \Psi_i^{l_2} \backslash \Gamma_2^{l_4}} \mathcal{P}_d^{ij_8} \label{group1_03} \hspace{0.5cm} 
\eeqn
where $H_i$ denotes the number of channels in $\mathcal{S}_i^{\text{com}}$. The quantity in (\ref{group1_00}) is the probability that
all available channels in $\mathcal{S}_i$ (if any) are overlooked by user $i$ due to false alarms. Therefore, user $i$ does not access
any channels in $\mathcal{S}_i$. The quantity in (\ref{group1_01}) describes the
probability that there are $k_2$ actually available channels in $\mathcal{S}_i^{\text{com}}$ and
 $\Psi_i^{l_2}$ denotes such a typical set with $k_2$ available channels.
The quantity in (\ref{group1_02}) describes the
probability that user $i$ correctly detects $k_3$ channels out of $k_2$ available channels.
The last quantity in (\ref{group1_03}) excluding the factor ${1}/{(k_3+k_4)}$
denotes the probability that user $i$ mis-detects $k_4$ channels among the remaining $H_i-k_2$ busy channels in $\mathcal{S}_i^{\text{com}}$.
Finally, the factor  ${1}/{(k_3+k_4)}$ is the probability that user $i$ correctly chooses one available channels in $\mathcal{S}_i^{\text{com}}$
for transmission out of $k_3+k_4$ channels which are indicated as being available by sensing.
\beqn
\Phi_1(A_1) = \sum \limits_{c_1=1}^{C_{\mathcal{MS}_j}^{A_1}} \prod \limits_{m_1 \in  \Omega_{c_1}^{(1)}  } \left(\mathcal{P}_{\text{idle}}^{m_1j} \left(1-\prod \limits_{l \in \mathcal{S}_{m_1}}\mathcal{P}_{\text{busy}}^{m_1 l}\right)\right).   \label{group1_I}
\eeqn
In (\ref{group1_I}), we consider all possible subsets of users of size $A_1$ that belongs to Group I (there are $C_{\mathcal{MS}_j}^{A_1}$ such subsets).
Each term inside the sum represents the probability of the corresponding event whose set of $A_1$ users is denoted by $\Omega_{c_1}^{(1)}$.
\beqn
\Phi_2(A_2) = \sum \limits_{c_2=1}^{C_{\mathcal{MS}_j-A_1}^{A_2}} \prod \limits_{m_2 \in \Omega_{c_2}^{(2)}  }  \mathcal{P}_{\text{busy}}^{m_2j}. \label{group1_II} 
\eeqn
In (\ref{group1_II}), we capture the probability that channel $j$ is indicated as not being available by sensing for $A_2$ users in group II. Possible sets of
these users are denoted by $\Omega_{c_2}^{(2)}$. 
\beqn
\Phi_3(A_3) = \sum \limits_{c_3=1}^{C_{\mathcal{MS}_j-A_1-A_2}^{A_3}} \prod \limits_{m_3 \in \Omega_{c_3}^{(3)}  } \left(\mathcal{P}_{\text{idle}}^{m_3j} \prod_{l_3 \in \mathcal{S}_{m_3}} \mathcal{P}_{\text{busy}}^{m_3l_3}\right)  \label{1group1_III} \hspace{3.0cm} \\ 
\hspace{0cm} \times   \left[  \sum \limits_{n=0}^{\beta} \sum_{q=1}^{C_{\beta}^n} 
\prod_{h_1 \in \mathcal{S}^{\text{com},q}_{j,m_3} }  \mathcal{P}_{\text{idle}}^{m_3 h_1} 
\prod_{h_2 \in \overline{\mathcal{S}}^{\text{com},q}_{j,m_3} }  \mathcal{P}_{\text{busy}}^{m_3 h_2}  
  \left( 1- \frac{1}{ n+1} \right) \right]. \label{2group1_III} 
\eeqn
For each term in (\ref{1group1_III}) we consider different possible subsets of $A_3$ users, which are denoted by $\Omega_{c_3}^{(3)}$.
Then, each term in (\ref{1group1_III})  represents the probability that channel $j$ is indicated as available by sensing for each user $m_3 \in \Omega_{c_3}^{(3)}$ 
while all channels in  $\mathcal{S}_{m_3}$ are indicated as not available by sensing. 
In (\ref{2group1_III}), we consider all possible sensing outcomes for channels in $\mathcal{S}_{m_3}^{\text{com}}$ performed by user
 $m_3 \in \Omega_{c_3}^{(3)}$.
In addition, let $\mathcal{S}^{\text{com}}_{j,m_3} = \mathcal{S}_{m_3}^{\text{com}} \backslash \left\{ j \right\}$  
and $\beta = |\mathcal{S}^{\text{com}}_{j,m_3}|$.
Then, in (\ref{2group1_III}) we consider all possible scenarios in which $n$ channels in $\mathcal{S}^{\text{com}}_{j,m_3}$ are indicated as
available by sensing; and user $m_3$ chooses a channel different from channel $j$ for transmission (with probability $\left( 1- \frac{1}{ n+1} \right)$) where
$\mathcal{S}^{\text{com}}_{j,m_3} = \mathcal{S}^{\text{com},q}_{j,m_3} \cup \overline{\mathcal{S}}^{\text{com},q}_{j,m_3} $ and
$\mathcal{S}^{\text{com},q}_{j,m_3} \cap \overline{\mathcal{S}}^{\text{com},q}_{j,m_3} = \emptyset$.
\beqn
\Phi_4(A_4) =  \left(\frac{1}{1+ A_4}\right) \prod \limits_{m_4 \in \Omega^{(4)} } \left(\mathcal{P}_{\text{idle}}^{m_4j} \prod_{l_4 \in \mathcal{S}_{m_4}} \mathcal{P}_{\text{busy}}^{m_4l_4}\right)  \label{1group1_IV}   \hspace{3cm}\\
\times \left[ \sum \limits_{m=0}^{\gamma} \sum_{q=1}^{C_{\gamma}^{m}}   
\prod_{h_1 \in \mathcal{S}^{\text{com},q}_{j,m_4} }  \mathcal{P}_{\text{idle}}^{m_4 h_1} 
\prod_{h_2 \in \overline{\mathcal{S}}^{\text{com},q}_{j,m_4} }  \mathcal{P}_{\text{busy}}^{m_4 h_2}  
\left( \frac{1}{ m+1} \right)  \right].
 \label{2group1_IV} 
\eeqn
The sensing outcomes captured in (\ref{1group1_IV}) and (\ref{2group1_IV}) are similar to those in (\ref{1group1_III}) and (\ref{2group1_III}).
However, given three sets of $A_1$, $A_2$, and $A_3$ users, the set $\Omega^{(4)}$ can be determined whose size is $|\Omega^{(4)}| = A_4$. 
Here, $\gamma$ denotes cardinality of the set $\mathcal{S}_{j,m_4}^{\text{com}} = \mathcal{S}_{m_4}^{\text{com}} \backslash \left\{ j \right\}$.
Other sets are similar to those in (\ref{1group1_III}) and (\ref{2group1_III}). However, all users in $\Omega^{(4)}$ choose channel $j$ for transmission
in this case. Therefore, user $i$ wins the contention with probability $1/(1+A_4)$ and its achievable throughput is $(1-\delta)/(1+A_4)$.

}
\end{itemize}

Summarize all considered cases, the throughput achieved by user $i$ is written as
\beqn
T_i = T_i \left\{\text{Case 1} \right\} + T_i \left\{ \text{Case 3} \right\} .
\eeqn
And the total throughput $\mathcal{T}$ can be calculated by summing the throughputs of all secondary users.

\subsection{Congestion of Control Channel}

Under our design, contention on the control channel is mild if the number of channels $N$ is relatively large compared to the number of SUs $M$.
In particular, there is no need to employ a MAC protocol if we have $N>>M$ since distinct sets of channels can be allocated for SUs
by using Algorithm 1. In contrast, if the number of channels $N$ is small compared to the
number of SUs $M$ then the control channel may experience congestion due to excessive control message exchanges. The congestion
of the control channel in such scenarios can be alleviated if we allow RTS/CTS messages to be exchanged in parallel on several channels 
(i.e., multiple rendezvous \cite{mo08}). Design of such a MAC protocol in the cognitive radio setting requires extra care compared
to traditional multi-channel settings since PUs must be protected. 

We describe potential design of a multiple-rendezvous
MAC protocol in the following using similar ideas of a multi-channel MAC protocol (McMAC) 
in \cite{mo08}, \cite{su08}. We assume that each SU hops through all channels by following a particular hopping pattern, which corresponds to 
a unique seed \cite{mo08}. In addition, each SU puts its seed in every packets so that neighboring SUs can learn its hopping pattern.
The same cycle structure as being described in Section V.A is employed here. Suppose SU A wishes to transmit data SU B in a particular
cycle. Then, SU A turns to the current channel of B and senses this channel as well as its assigned channels in $\mathcal{S}_{\sf AB}^{\sf tot}$,
which is the set of allocated channels for link $AB$. If SU A's sensing outcomes indicate that the current channel of SU B is available then SU A sends RTS/CTS messages with SU B containing a chosen available communication channel. Otherwise, SU A waits until the next cycle to perform sensing and contention again. If the handshake is successful, SU A transmits data to SU B on the chosen channel in the data phase. Upon completing
data transmission, both SUs A and B return to their home hopping patterns.

It is worth noting that the throughput analysis performed in Section VI.A and Section VII.B is still valid here except that 
we have to derive the protocol overhead and choose an appropriate contention window under this new design. In general, collisions 
among SUs are less frequent under
a multiple-rendezvous MAC protocol since contentions can occur in parallel on different channels. As suggested by  \cite{mo08},
it would not be possible to design a multi-channel MAC protocol that can work efficiently in all different scenarios.  
Discussions on different potential designs of a multi-channel MAC protocol and their corresponding pros/cons can be found in \cite{mo08}
and the references therein. We would like to emphasize that the focus of this
paper is on the channel assignment issue; therefore, consideration of alternative designs of a MAC protocol is beyond its scope.

\section{Numerical Results}
\label{Results}

\textbf{We present numerical results to illustrate the throughput performance of the proposed channel assignment algorithms. 
To obtain the results, the probabilities $p_{i,j}$ are randomly realized in the interval [0.7, 0.9]. 
We choose the length of control packets as follows: RTS including PHY header 288 bits, CTS including PHY header 240 bits,
which correspond to $t_{\sf RTS}$ = 48$\mu s$, $t_{\sf CTS}$ = 40 $\mu s$ for transmission rate of 6Mbps, which is
the basic rate of 802.11a/g standards. Other parameters are chosen as follows: cycle time 
$T_{\sf cycle} = 3 ms$; $\theta = 20$ $\mu s$, $t_{\sf SIFS}$ = 28 $\mu s$,
target collision probability $\epsilon_P$ = 0.03; $t_{\sf SEN}$ and  $t_{\sf SYN}$ are assumed to be negligible so they are ignored.
Note that these values of $\theta$ and $t_{\sf SIFS}$ are typical (interest readers can refer to Tables I and II
for \cite{bian00} for related information). The value of cycle time $T_{\sf cycle}$ is chosen based on the fact
that practical cognitive systems such as those operating on the TV bands standardized in the 802.22 standard requires 
maximum evacuation time of a few milliseconds \cite{ko10}.}

\subsection{ MAC Protocol Configuration}

\begin{figure}[!t]
\centering
\includegraphics[width=90mm]{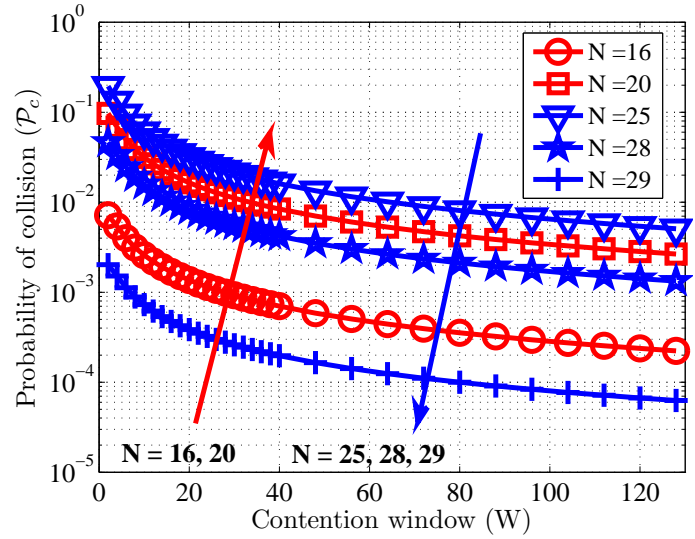}
\caption{Collision probability versus the contention window (for $M$ = 15).}
\label{het_P_M15_Aware_Blind}
\end{figure}

We first investigate interactions between MAC protocol parameters and the achievable throughput performance.
In particular, we plot the average probability of the first collision, which is derived in Section \ref{ConWinCal} versus 
contention window in Fig.~\ref{het_P_M15_Aware_Blind} when Algorithm 2 is used for channel assignment. This figure shows that 
the collision probability first increases then decreases with $N$. This can be
interpreted as follows. When $N$ is relatively small, Algorithm 2 tends to allow
more overlapping channel assignments for increasing number of channels. 
However, more overlapping channel assignments increase the contention level because
more users may want to exploit same channels, which results in larger collision probability.
As $N$ is sufficiently large, a few overlapping channel assignments is needed to achieve
the maximum throughput. Therefore, collision probability decreases with $N$.

\begin{figure}[!t]
\centering
\includegraphics[width=90mm]{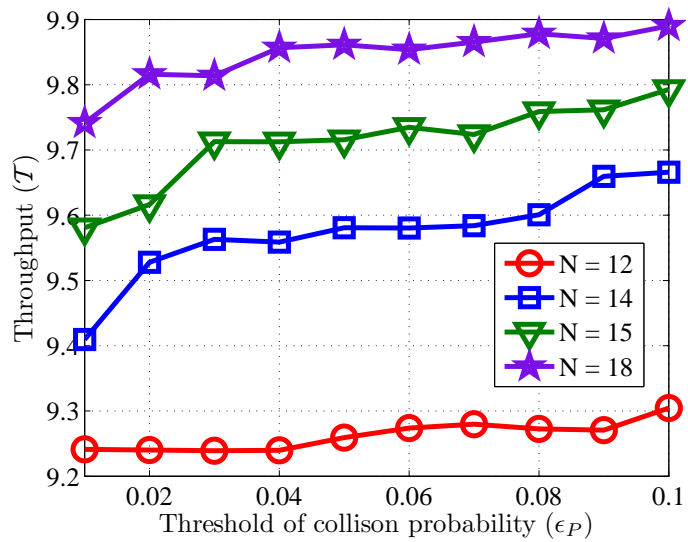}
\caption{Throughput versus target collision probability (for $M$ = 10)}
\label{T_Pc_M10_N121518}
\end{figure}

We now consider the impact of target collision probability $\epsilon_P$ on the total network throughput,
which is derived in Section \ref{tputana}. Recall that in this analysis collision probability
is not taken into account, which is shown to have negligible errors in Proposition 2. Specifically, we plot
the total network throughput versus $\epsilon_P$ for $M$ = 10 and different values of $N$ in Fig.~\ref{T_Pc_M10_N121518}.
This figure shows that the total throughput slightly increases with $\epsilon_P$. However, the increase is quite marginal
as $\epsilon_P \geq 0.03$. In fact, the required contention window $W$ given in (\ref{Window}) decreases with increasing
$\epsilon_P$ (as can be observed from Fig.~\ref{het_P_M15_Aware_Blind}), which leads to decreasing MAC protocol
overhead $\delta(W)$ as confirmed by (\ref{overhead}) and therefore the increase in the total network throughput.
Moreover, the total throughput may degrade with increasing $\epsilon_P$ because of the increasing number of collisions.
Therefore, choosing $\epsilon_P=0.03$ would be reasonable to balance between throughput gain due to moderate
MAC protocol overhead and throughput loss due to contention collision. We will illustrate the throughput performance
achieved by our proposed algorithms for this value of target collision probability in the next sub-section.

\subsection{Comparisons of Proposed Algorithms versus Optimal Algorithms}

\begin{figure}[!t]
\centering
\includegraphics[width=90mm]{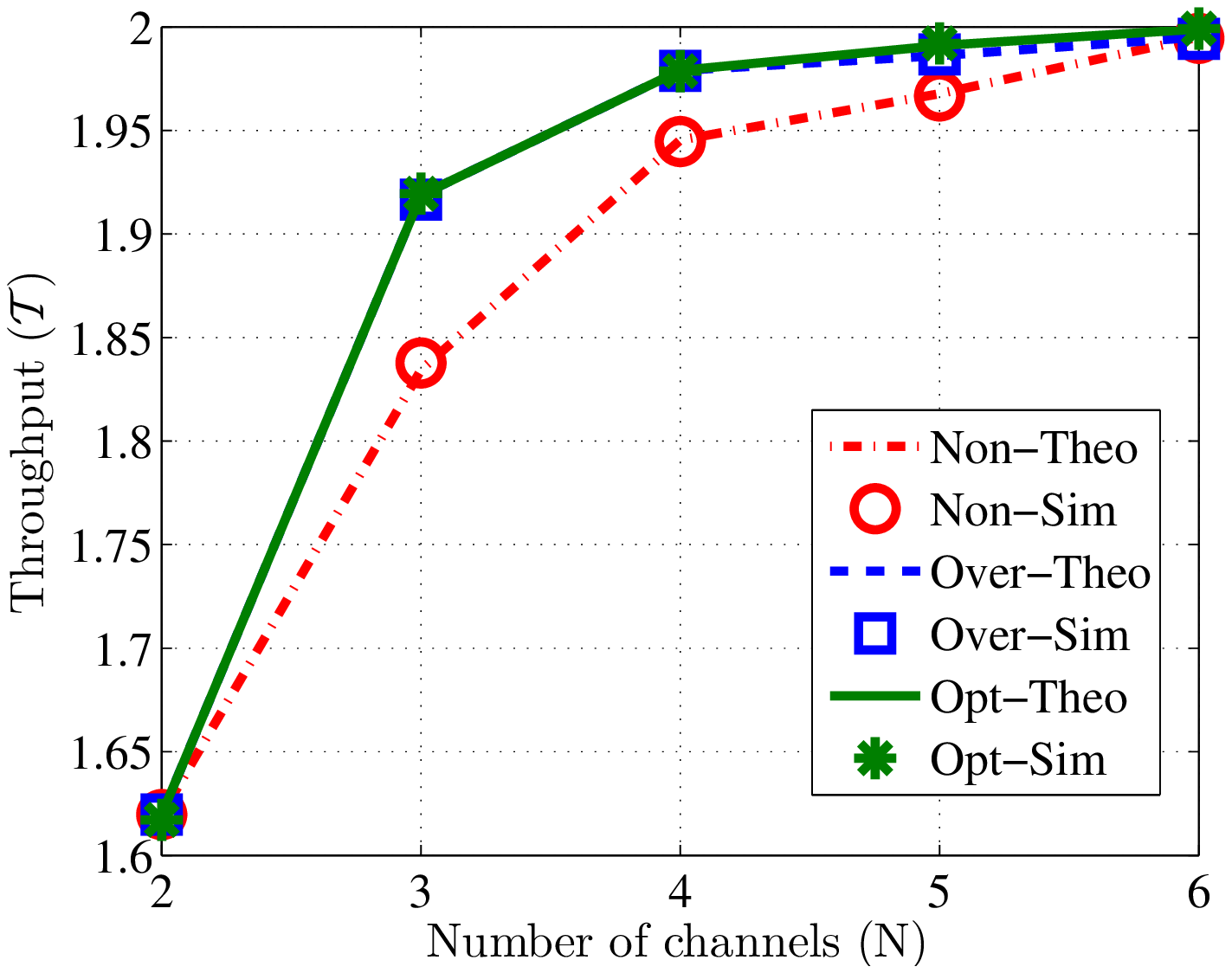}
\caption{Throughput versus the number of channels (for $M $ = 2, Theo: Theory, Sim: Simulation, Over: Overlapping, Non: Non-overlapping, Opt: Optimal assignment).}
\label{M_2_Opt_Comparison}
\end{figure}

\begin{figure}[!t]
\centering
\includegraphics[width=90mm]{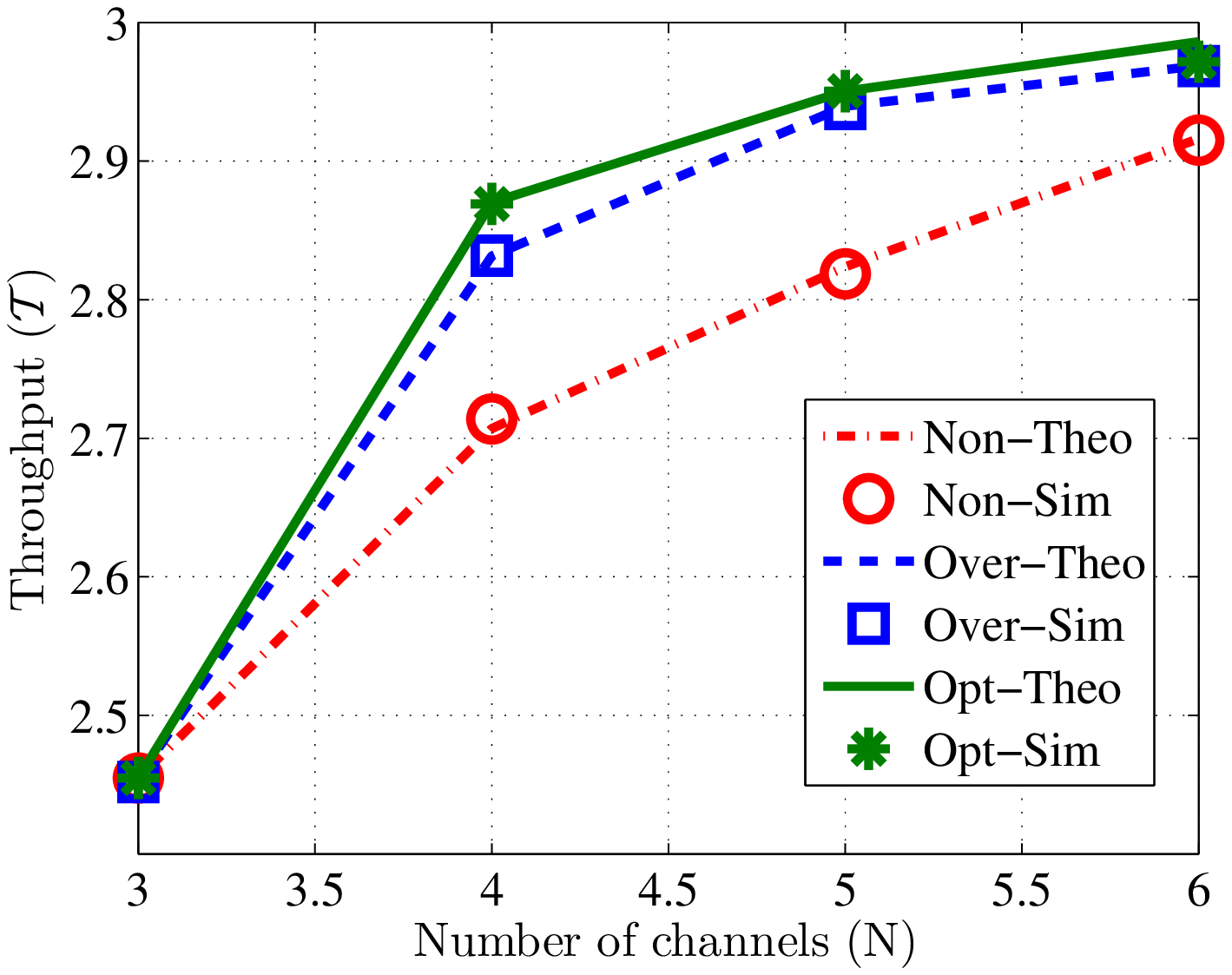}
\caption{Throughput versus the number of channels (for $M $ = 3, Theo: Theory, Sim: Simulation, Over: Overlapping, Non: Non-overlapping, Opt: Optimal assignment).}
\label{M_3_Opt_Comparison}
\end{figure}

\textbf{We demonstrate the efficacy of the proposed algorithms by comparing their throughput performances with those obtained
by the optimal brute-force search algorithms for small values of $M$ and $N$. Numerical results are presented for both 
throughput-maximization and max-min fair objectives. In Figs.~ \ref{M_2_Opt_Comparison} and \ref{M_3_Opt_Comparison}, 
we compare the throughputs of the proposed and optimal
algorithms for $M = 2$ and $M = 3$ under the throughput-maximization objective. These figures confirm that Algorithm 2
achieves very close to the optimal solutions for both values of $M$. }

\begin{figure}[!t]
\centering
\includegraphics[width=90mm]{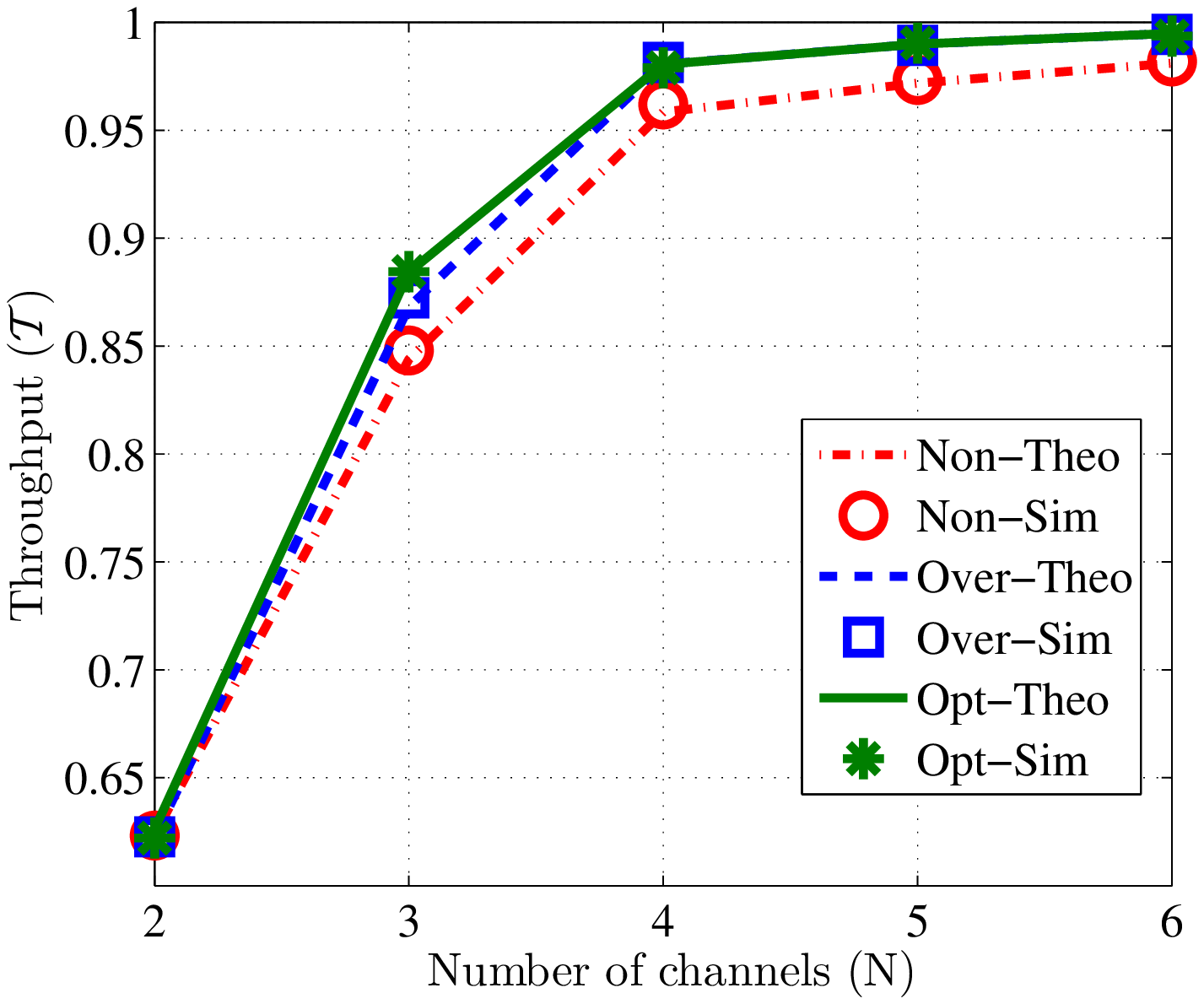}
\caption{Minimum throughput versus the number of channels (for $ M$ = 2, Theo: Theory, Sim: Simulation, Over: Overlapping, Non: Non-overlapping, Opt: Optimal assignment).}
\label{M_2_fairness_optimal_compare}
\end{figure}

\begin{figure}[!t]
\centering
\includegraphics[width=90mm]{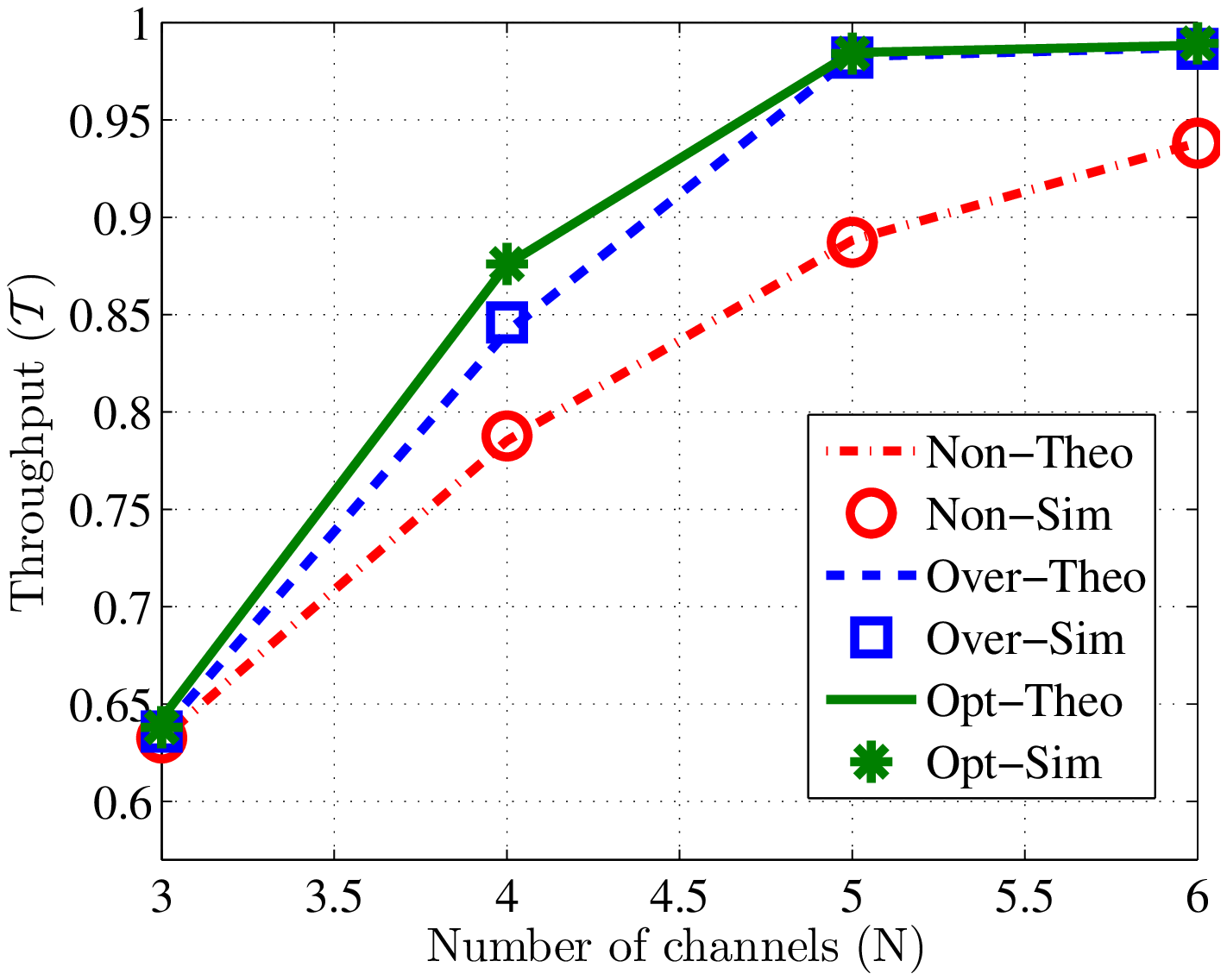}
\caption{Minimum throughput versus the number of channels (for $M$ = 3, Theo: Theory, Sim: Simulation, Over: Overlapping, Non: Non-overlapping, Opt: Optimal assignment).}
\label{M_3_fairness_optimal_compare}
\end{figure}

\textbf{In Figs.~ \ref{M_2_fairness_optimal_compare}, \ref{M_3_fairness_optimal_compare}, we plot the throughputs achieved by
our proposed algorithm and the optimal algorithm for $M = 2$ and $M = 3$ under the max-min fair objective.
Again Algorithm 4 achieves throughput very close to the optimal throughput under this fair objective.
These results are very positive given that Algorithm 2 and Algorithm 4 have much lower complexity than those
of the optimal brute-force search algorithms. In addition, analytical results match simulation results very well
and non-overlapping channel assignment algorithms achieve noticeably lower throughputs than those by their overlapping counterparts.}

\subsection{Throughput Performance of Proposed Algorithms}
 
\begin{figure}[!t]
\centering
\includegraphics[width=90mm]{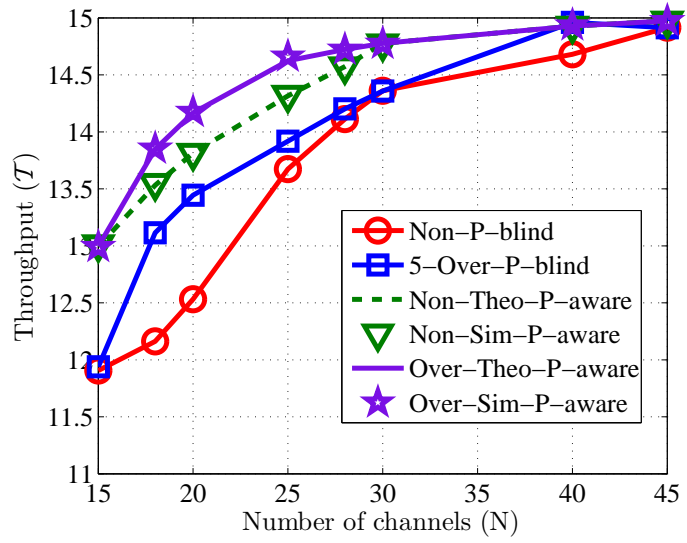}
\caption{Throughput versus the number of channels (for $M$ = 15, Theo: Theory, Sim: Simulation, Over: Overlapping, Non: Non-overlapping, 5-Over: 5-user sharing Overlapping )}
 \label{het_T_M10_Aware_Blind1}
\end{figure}

We illustrate the total throughput $\mathcal{T}$ versus the number of channels obtained by both Algorithm 1 and Algorithm 2 where each point is obtained by averaging the throughput over 30 different realizations of $p_{i,j}$ in Fig.~\ref{het_T_M10_Aware_Blind1}. Throughput curves 
due to Algorithms 1 and 2 are indicated as ``P-ware''  in this figure. In addition, for the comparison purposes, we also show the throughput performance achieved by ``P-blind'' algorithms, which simply allocate channels to users in a round-robin manner without exploiting the heterogeneity
of $p_{i,j}$ (i.e., multiuser diversity gain). For P-blind algorithms, we show the performance of both non-overlapping and overlapping channel assignment algorithms. Here, the overlapping P-blind algorithm allows at most five users to share one particular channel. We have observed through
numerical studies that allowing more users sharing one channel cannot achieve better throughput performance because of the excessive MAC protocol overhead.

As can be seen in Fig.~\ref{het_T_M10_Aware_Blind1}, analytical and simulation results achieved by  both proposed algorithms match each other very well.
This validates the accuracy of our throughput analytical model developed in Section \ref{tputana}. 
It also indicates  that the total throughput reaches the maximum value, which is equal to $M$ = 15 as the 
number of channels becomes sufficiently large for both Algorithms 1 and 2. This confirms the result stated in Proposition 1. In addition, Algorithm 2 achieves significantly larger throughput than Algorithm 1 for low or moderate values of $N$. This performance gain comes from the multiuser diversity gain, which arises due to the spatial dependence of white spaces. For large $N$ (i.e., more than twice the number of users $M$), the negative impact
of MAC protocol overhead prevents Algorithm 2 from performing overlapped channel assignments. Therefore, both Algorithms 1 and 2 achieve similar throughput performance.

\begin{figure}[!t]
\centering
\includegraphics[width=90mm]{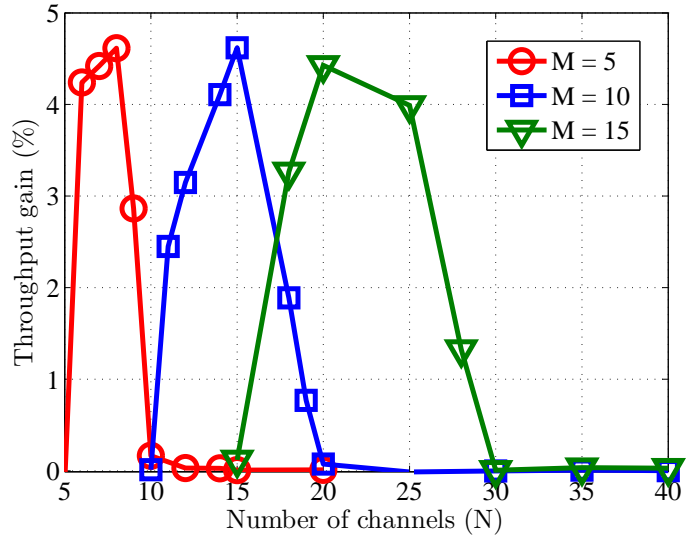}
\caption{Throughput gain between Algorithm 2 and Algorithm 1 versus the number of channels}
\label{het_G_M_351015}
\end{figure}

\begin{figure}[!t]
\centering
\includegraphics[width=90mm]{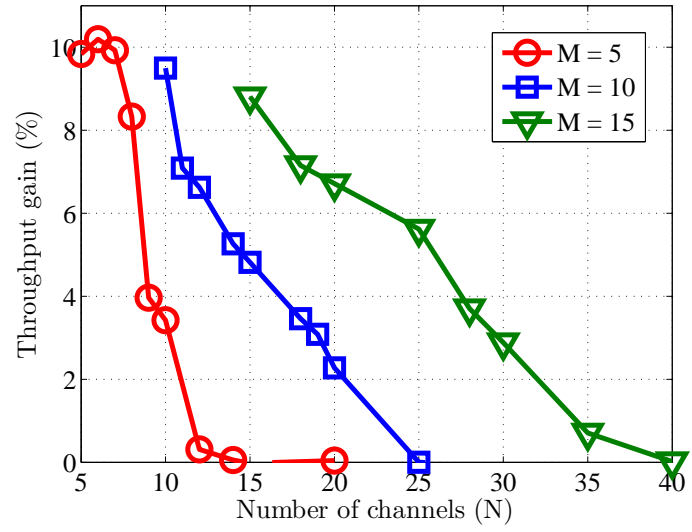}
\caption{Throughput gain between Algorithm 2 and P-blind 5-user sharing versus the number of channels }
\label{Paware_Pblind_M_51015_G}
\end{figure}

Fig.~\ref{het_T_M10_Aware_Blind1} also indicates that both proposed algorithms outperform the round-robin channel assignment counterparts. In particular, Algorithm 1 improves the total throughput significantly compared to the round-robin algorithm under non-overlapping channel assignments. For the overlapping channel assignment schemes, we show the throughput performance of the round-robin assignment algorithms when 5 users are allowed to share one channel (denoted as 5-user sharing in the figure). Although this achieves larger throughput for the round-robin algorithm, it still performs worse compared to the proposed algorithms. 
Moreover, we demonstrate the throughput gain due to Algorithm 2 compared to Algorithm 1 for different values
of $N$ and $M$ in Fig.~\ref{het_G_M_351015}. This figure shows that performance gains up to $5\%$ can be achieved 
when the number of channels is small or moderate. Also, Fig.~\ref{Paware_Pblind_M_51015_G} presents the throughput gain due to Algorithm 2 versus the P-blind algorithm with 5-user sharing. It can be observed that a significant throughput gain of up to 10\% can be achieved for these investigated scenarios.

\begin{figure}[!t]
\centering
\includegraphics[width=90mm]{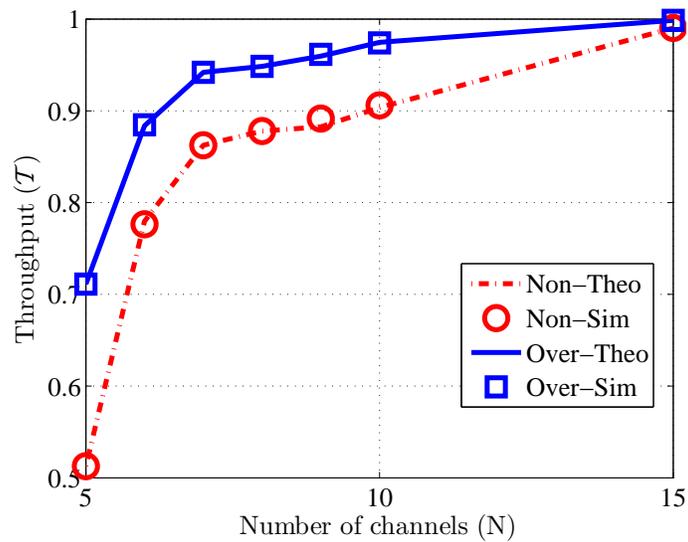}
\caption{Minimum throughput versus the number of channels (for $M$ = 5, Theo: Theory, Sim: Simulation, Over: Overlapping, Non: Non-overlapping).}
\label{M5_fairness}
\end{figure}

\begin{figure}[!t]
\centering
\includegraphics[width=90mm]{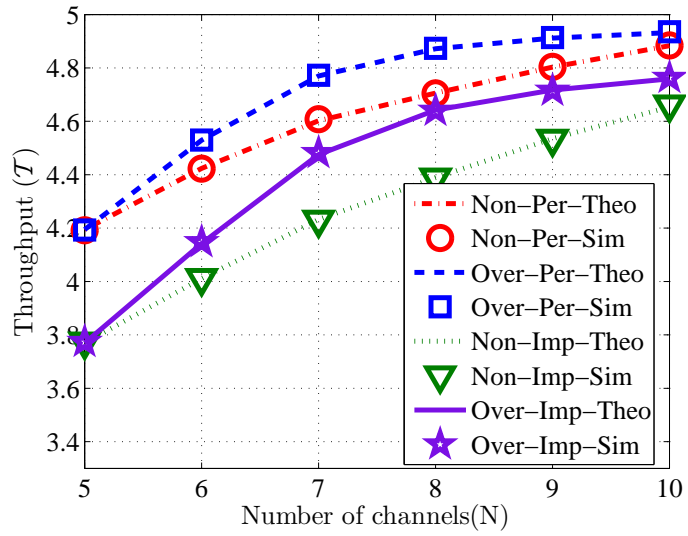}
\caption{Throughput versus the number of channels (for $M = 5, \mathcal{P}_f^{ij} = \left[0.1, 0.15\right], \mathcal{P}_d^{ij} = 0.9$, Theo: Theory, Sim: Simulation, Over: Overlapping, Non: Non-overlapping, Per: Perfect, Imp: Imperfect).}
\label{M5_Pf_10_tau50}
\end{figure}

\textbf{Fig.~ \ref{M5_fairness} illustrates the throughput of Algorithms 3 and 4 where $p_{ij}$ are chosen in the range of $\left[0.5,0.9\right]$.
It can be observed that the overlapping channel algorithm improves the throughput performance compared to the non-overlapping
counterpart in terms of the minimum throughput. Finally, we plot the throughputs achieved by Algorithms 1 and 2 under perfect
and imperfect spectrum sensing for $M = 5$ in Fig.~\ref{M5_Pf_10_tau50}. This figure shows that sensing errors can significantly
degrade the throughput performance of SUs. In addition, the presented results validate the throughput analytical model described
in Section VII.B.}

\vspace{0.2cm}
\section{Conclusion}
\label{conclusion} 

We have considered a channel assignment problem for cognitive radio networks
with hardware-constrained secondary users in this paper.
We have presented the optimal brute-force search algorithm and analyzed its complexity.
To resolve the high complexity of the optimal search,
we have developed two channel assignment algorithms for throughput maximization. The first algorithm performs non-overlapping channel assignment 
for secondary users, which was shown to achieve optimality if the number of channels is sufficiently large. In the
secondary algorithm, we have allowed overlapping channel assignments and designed a MAC protocol to resolve channel access
contention when different users attempt to exploit the same available channel. In addition, we have developed an analytical
model to analyze the saturation throughput achieved by Algorithm 2. We have presented several potential extensions
including design of max-min fair channel assignment algorithms, throughput analysis considering imperfect spectrum sensing, 
and alternative MAC protocol design to relieve congestion of the control channel.
 We have validated our results via
numerical studies and demonstrated significant throughput gains of the overlapping channel assignment algorithm
compared to the non-overlapping and round-robin channel assignment counterparts in different network settings.



\renewcommand{\baselinestretch}{1.6}

\bibliographystyle{IEEEtran}



\newpage

\renewcommand{\baselinestretch}{1.0}

\begin{algorithm}[h]
\caption{\textsc{Overlapping Channel Assignment}}
\label{mainalg}
\begin{algorithmic}[1]

\STATE Initialize the sets of allocated channels for all users  $\mathcal{S}_i := \emptyset$ for $i=1, 2,\ldots , M$ and $\delta_0$

\STATE Run Algorithm 1 to obtain non-overlapping channel assignment solution.

\STATE Let the group of channels shared by $l$ users be $\mathcal{G}_l$
and $\mathcal{U}_j$ be the set of users sharing channel $j$ and set $\mathcal{U}_j^\text{temp} := \mathcal{U}_j, \: \forall j = 1, 2, \ldots, N$.

\STATE continue := 1; $h$ = 1; updoverhead := 0

\WHILE {$\text{continue}  = 1$}

\STATE Find the group of channels shared by $h$ users, $\mathcal{G}_{h}$

\FOR {$j=1$ to $\left|\mathcal{G}_{h}\right|$}

\FOR {$l=1$ to $M$}

\IF{$l \in \mathcal{U}_j $}

\STATE $\Delta T_{l}^{h, \text{est}} (j) = 0$

\ELSE

\STATE User $l$ calculates $\Delta T_l^{h, \text{est}} (j)$ assuming channel $j$ is allocated to user $l$

\ENDIF

\ENDFOR

\STATE ${l^*_j} = \argmax_{l}  \Delta T_l^{h, \text{est}} (j) $.

\ENDFOR

\STATE ${j_{l^*}^*} = \argmax_j  \Delta T_{l^*_j}^{h, \text{est}} (j) $.

\IF{ $\Delta T_{l^*}^{h, \text{est}} (j_{l^*}^*) \leq \epsilon$ and \text{updoverhead} = 1}

\STATE Set: \text{continue} := 0

\STATE Go to step 35

\ENDIF

%
%

\IF{ $\Delta T_{l^*}^{h, \text{est}} (j_{l^*}^*) > \epsilon$}

\STATE Temporarily assign channel $j_{l^*}^*$ to user $l^*$, i.e., 
 update $\mathcal{U}_{j_{l^*}^*}^{\text{temp}} = \mathcal{U}_{j_{l^*}^*}  \cup \left\{ l^* \right\} $;

\STATE Calculate $W$ and $\delta$ with  $\mathcal{U}_{j_{l^*}^*}^{\text{temp}}$ by using methods in Sections \ref{ConWinCal} and \ref{Overcal}, respectively. 

\IF{ $\left| \delta - \delta_0 \right| > \epsilon_{\delta}$}

\STATE Set: \text{updoverhead} := 1

\STATE Return Step 7 using the updated $\delta_0 =\delta $

\ELSE 

\STATE Update  $\mathcal{U}_{j_{l^*}^*} := \mathcal{U}_{j_{l^*}^*}^{\text{temp}}$ 
 (i.e., assign channel $j_{l^*}^*$ to user $l^*$), calculate $W$ and $\delta_0$ with $\mathcal{U}_{j_{l^*}^*}$, and
 update $\mathcal{G}_{h}$
 
 \STATE Update: \text{updoverhead} := 0

\ENDIF

\ENDIF

\STATE Return Step 7 

\STATE $h=h+1$

\ENDWHILE

\end{algorithmic}
\end{algorithm}

\begin{algorithm}[h]
\caption{\textsc{Fair Non-Overlapping Channel Assignment}}
\label{mainalgFN}
\begin{algorithmic}[1]

\STATE Initialize SU $i$'s set of available channels, ${\mathcal{S}_i^a} := \left\{ {1,2, \ldots ,N} \right\}$ and  $\mathcal{S}_i := \emptyset$ for $i=1, 2,\ldots , M$ where $\mathcal{S}_i$ denotes the set of channels assigned for SU $i$.

\STATE $\text{continue} := 1$

\WHILE {$\text{continue} = 1$}

\STATE Find the set of users who currently have minimum throughput $\mathcal{S}^{\text{min}} = \mathop {\argmin} \limits_i  T_i^b $

where $\mathcal{S}^{\text{min}} = \left\{i_1, \ldots, i_m\right\} \subset \left\{1,\ldots,M\right\}$ is the set of minimum-throughput SUs.

\IF {$\mathop {\mathcal{OR}} \limits_{i_l \in \mathcal{S}^{\text{min}}} \left(\mathcal{S}_{i_l}^a \neq \emptyset\right)$}

\STATE For each SU $i_l \in \mathcal{S}^{\text{min}}$ and channel $j_{i_l} \in \mathcal{S}_{i_l}^a$, find $\Delta T_{i_l}(j_{i_l}) = T_{i_l}^a - T_{i_l}^b$

where $T_{i_l}^a$ and $T_{i_l}^b$ are the throughputs after and before assigning channel $j_{i_l}$; and we set $\Delta T_{i_l} = 0$ if $\mathcal{S}_{i_l}^a = \emptyset$.

\STATE $\left\{i_{l}^{*}, j_{i_{l}^*}^* \right\}= \mathop {\argmax }\limits_{i_l \in \mathcal{S}^{\text{min}}, j_{i_l} \in {\mathcal{S}_{i_l}^a}} \: \Delta {T_{i_l}}(j_{i_l})$

\STATE Assign channel $j_{i_{l}^*}^*$ to SU $i_{l}^*$.

\STATE Update $\mathcal{S}_{i_{l}^*} = \mathcal{S}_{i_{l}^*} \cup j_{i_{l}^*}^*$ and $\mathcal{S}_k^a = \mathcal{S}_k^a \backslash j_{i_{l}^*}^*$ for all $k \in \left\{1,\ldots,M\right\}$.

\ELSE

\STATE Set $\text{continue} := 0$

\ENDIF

\ENDWHILE

\end{algorithmic}
\end{algorithm}

\begin{algorithm}[h]
\caption{\textsc{Fair Overlapping Channel Assignment}}
\label{mainalgFO1}
\begin{algorithmic}[1]


\STATE Run Algorithm 3 and obtain the sets $\mathcal{S}_i$ for all SU $i$. Initialize $\mathcal{S}_i^{\text{com}}=\emptyset$ for $i$.

\STATE continue := 1.

\WHILE {$\text{continue}  = 1$}

\STATE Find  $i^* = \mathop {\argmin} \limits_{i \in \left\{1,\ldots,M \right\}} T_i^b$ and $T_{\text{min}} = T_{i^*}^b$ where ties
are broken randomly.

\STATE $\mathcal{S}_{i^*}^{\text{Sep}} = \mathop \cup \limits_{i, i \neq i^*}  \mathcal{S}_{i} $, $\mathcal{S}_{i^*}^{\text{Uni}} = \mathop \cup \limits_i  \mathcal{S}_{i}^{\text{com}} \backslash \mathcal{S}_{i^*}^{\text{com}}$.

\STATE Run Algorithm 5.


\IF {$\mathop \mathcal{OR} \limits_i  \mathcal{S}_i^{\text{com},\text{temp}} \neq \emptyset$}

\STATE Assign $\mathcal{S}_i^{\text{com}} = \mathcal{S}_i^{\text{com},\text{temp}}$ and $\mathcal{S}_i = \mathcal{S}_i^{\text{temp}}$.

\ELSE

\STATE Set $\text{continue} := 0$.

\ENDIF

\ENDWHILE

\end{algorithmic}
\end{algorithm}

\begin{algorithm}[h]
\caption{\textsc{Searching Potential Channel Assignment}}
\label{mainalgFO2}
\begin{algorithmic}[1]


\STATE --- \textit{Search potential channel assignment from separate sets} ---
\vspace{0.2cm}

\FOR {$j \in \mathcal{S}_{i*}^{\text{Sep}}$}

\STATE Find SU $i'$ where $j \in S_{i'}$. Let $n_c = M - 2$. 

\FOR {$l = 0$ to $n_c$}

\FOR {$k = 1$ to $C_{n_c}^l$}

\STATE Find $T_{i^*}^a $, $T_{i'}^a $, and $T_m^a\left|_{m \in \mathcal{U}^l_j}\right.$, where $\mathcal{U}^l_j$ is the set of $l$ new SUs sharing channel $j$.

\IF { $\min \left(T_{i^*}^a , T_m^a\left|_{m \in \mathcal{U}^l_j}\right., T_{i'}^a \right) > T_{\text{min}}$ }

\STATE - Temporarily assign channel $j$ to SUs $i^*$, $i'$ and all SUs $m$: $\mathcal{S}_{i^*}^{\text{com},\text{temp}} = \mathcal{S}_{i^*}^{\text{com}} \cup j$, $\mathcal{S}_{i'}^{\text{com},\text{temp}} = \mathcal{S}_{i'}^{\text{com}} \cup j$, $\mathcal{S}_{i'}^{\text{temp}} = \mathcal{S}_{i'} \backslash j$ and $\mathcal{S}_m^{\text{com},\text{temp}} = \mathcal{S}_m^{\text{com}} \cup j$ . 

\STATE - Update $T_{\text{min}} = \min \left(T_{i^*}^a , T_m^a\left|_{m \in \mathcal{U}^l_j}\right. , T_{i'}^a \right)$. 

\STATE - Reset all temporary sets of other SUs to be empty.

\ENDIF

\ENDFOR

\ENDFOR

\ENDFOR

\vspace{0.4cm}

\STATE --- \textit{Search potential channel assignment from common sets} ---
\vspace{0.2cm}

\FOR {$j \in \mathcal{S}_{i^*}^{\text{Uni}}$}

\STATE Find the subset of SUs except SU $i^*$, $\mathcal{S}^{\text{Use}}$ who use channel $j$ as an overlapping channel.

\FOR {$l = 0$ to $M-1-\left|\mathcal{S}^{\text{Use}}\right|$}

\FOR {$k = 1$ to $C_{M-1-\left|\mathcal{S}^{\text{Use}}\right|}^l$}

\STATE Find $T_{i^*}^a $, $T_{i'}^a\left|_{i' \in \mathcal{S}^{\text{Use}}} \right.$, $T_m^a\left|_{m \in \mathcal{U}^l_j} \right.$, where $\mathcal{U}^l_j$ is the set of $l$ new SUs sharing channel $j$.

\IF { $\min \left(T_{i^*}^a , T_{i'}^a\left|_{i' \in \mathcal{S}^{\text{Use}}} \right., T_m^a\left|_{m \in \mathcal{U}^l_j} \right. \right) > T_{\text{min}}$}

\STATE - Temporarily assign channel $j$ to SU $i^*$, all SUs $i'$ and all SUs $m$: $\mathcal{S}_{i^*}^{\text{com},\text{temp}} = \mathcal{S}_{i^*}^{\text{com}} \cup j $, $\mathcal{S}_m^{\text{com},\text{temp}} = \mathcal{S}_m^{\text{com}} \cup j$. 

\STATE - Update $T_{\text{min}} = \min \left(T_{i^*}^a , T_{i'}^a\left|_{i' \in \mathcal{S}^{\text{Use}}} \right., T_m^a\left|_{m \in \mathcal{U}^l_j} \right. \right)$. 

\STATE - Reset all temporary sets of other SUs to be empty.

\ENDIF

\ENDFOR

\ENDFOR

\ENDFOR

\end{algorithmic}
\end{algorithm}

%
%
%
%
%
%
%
%
%
%
%
%
%
%
%

\end{document}